\documentclass[aps,pra,twocolumn,superscriptaddress]{revtex4-2}
\usepackage{float}
\usepackage{graphicx}
\usepackage{dsfont}
\usepackage[latin1]{inputenc}
\usepackage{version} 
\usepackage{color}
\usepackage{amssymb}
\usepackage{amsmath, scalerel}
\usepackage{braket}
\usepackage{bm}
\usepackage{upgreek}
\usepackage{colortbl}
\usepackage{graphicx}
\usepackage{dcolumn}
\usepackage{bm}
\usepackage{bigints}
\usepackage{xcolor}  
\usepackage[pagebackref=false]{hyperref} 
\usepackage{enumitem}
\usepackage{ulem}

\definecolor{jrp}{rgb}{1,0,0}

\definecolor{rim}{rgb}{0,1,0}

\definecolor{pr}{rgb}{0.7,0,0}

\definecolor{mjg}{rgb}{.08,.05,.8}

\definecolor{yyl}{rgb}{0.7,0.0,0.1}

\newcommand{\delete}[1]{{}} 
\colorlet{shadecolor}{gray!40}

\hypersetup{
    bookmarks=true,         
    unicode=false,          
    pdftoolbar=true,        
    pdfmenubar=true,        
    pdffitwindow=false,     
    pdfstartview={FitH},    
    pdftitle={Majorana_Main},    
    pdfauthor={X. Mi},     
    pdfkeywords={keyword1} {key2} {key3}, 
    pdfnewwindow=true,      
    colorlinks=true,       
    linkcolor=red,          
    citecolor=blue,        
    filecolor=magenta,      
    urlcolor=cyan,           
}

\makeatletter

\makeindex

\begin{document}
\title{Noise resilience of edge modes on a chain of superconducting qubits}

\newcommand{\xGoogle}{\affiliation{Google Research, Mountain View, CA, USA}}

\newcommand{\xGeneva}{\affiliation{Department of Theoretical Physics, University of Geneva, Quai Ernest-Ansermet 30, 1205 Geneva, Switzerland}}

\newcommand{\xUMass}{\affiliation{Department of Electrical and Computer Engineering, University of Massachusetts, Amherst, MA, USA}}

\newcommand{\xYale}{\affiliation{Department of Applied Physics, Yale University, New Haven, CT 06520, USA}}

\newcommand{\xCaltech}{\affiliation{Institute for Quantum Information and Matter, California Institute of Technology, Pasadena, CA, USA}}

\newcommand{\xUCR}{\affiliation{Department of Electrical and Computer Engineering, University of California, Riverside, CA, USA}}

\newcommand{\xUCSB}{\affiliation{Department of Physics, University of California, Santa Barbara, CA, USA}}

\author{X. Mi}\xGoogle
\author{M. Sonner}\xGeneva
\author{M. Y. Niu}\xGoogle
\author{K. W.~Lee}\xGoogle
\author{B. Foxen}\xGoogle
\author{R. Acharya}\xGoogle
\author{I. Aleiner}\xGoogle
\author{T. I.~Andersen}\xGoogle
\author{F. Arute}\xGoogle
\author{K. Arya}\xGoogle
% ...

\begin{abstract}
Inherent symmetry of a quantum system may protect its otherwise fragile states. Leveraging such protection requires testing its robustness against uncontrolled environmental interactions. Using 47 superconducting qubits, we implement the one-dimensional kicked Ising model which exhibits non-local Majorana edge modes (MEMs) with $\mathbb{Z}_2$ parity symmetry. Remarkably, we find that any multi-qubit Pauli operator overlapping with the MEMs exhibits a uniform late-time decay rate comparable to single-qubit relaxation rates, irrespective of its size or composition. This characteristic allows us to accurately reconstruct the exponentially localized spatial profiles of the MEMs. Furthermore, the MEMs are found to be resilient against certain symmetry-breaking noise owing to a prethermalization mechanism. Our work elucidates the complex interplay between noise and symmetry-protected edge modes in a solid-state environment.

\end{abstract}  

\maketitle

The symmetry of quantum systems can give rise to topologically distinct degenerate ground states. The quantum superposition of such states is in principle immune to dephasing, and an additional energy gap from the excited states further protects the ground states from energy decay. As such, symmetry-protected ground states may form decoherence-free subspaces \cite{Zanardi_PRL_1997,Lidar1998,Bacon2000,Kitaev2003} and are promising candidates for topological quantum computing \cite{Nayak_RMP_2008,Fowler2012}. An example model supporting symmetry-protected topological states is the Kitaev model of spinless fermions in a 1D wire~\cite{Kitaev_2001}. The $\mathbb{Z}_2$ parity symmetry of the model leads to a pair of degenerate ground states. The distinct parities of the two ground states protect them against local parity-preserving noise, such as potential fluctuations~\cite{ReadGreen}. The topological property of these degenerate ground states is commonly described by a pair of localized Majorana edge modes (MEMs) at the ends of the wire.

While the degree of symmetry protection in a closed quantum system is often understood, experimental quantum systems are invariably subject to physical noise sources that do not necessarily respect the underlying symmetry. In the context of MEMs, significant efforts have been directed toward experimentally realizing the Kitaev model, e.g. in nano-wires with spin-orbit interactions placed in the proximity of a superconductor \cite{LutchynMajorana, OregMajorana, MajoranaDelft1, RokhinsonMajorana, DasMajorana, Yazdani2014, MarcusMajorana2016,Frolov2021}. Here the underlying $\mathbb{Z}_2$ symmetry cannot be broken by local perturbations within a closed system. Nevertheless, theoretical results have widely suggested that MEMs remain susceptible to a variety of decoherence effects from their open solid-state environment \cite{Goldstein_PRB_2011_decay, Cheng_PRB_2012_Protection, Budich_PRB_2012_decay, Knapp_PRB_2018}. Experimental results have also established that the density of sub-gap quasiparticles is often orders of magnitude higher than predictions from simple thermal population arguments \cite{Falk_PRL_2014,Feldman2016,Serniak_PRL_2018, Hays_PRL_Parity_2018}. The incoherent processes involving these quasiparticles can change the parity of the ground state,  and consequently destroy the topological protection. These results highlight the importance of characterizing the extent of symmetry protection in realistic open-system environments.

\begin{figure*}[t!]
    \centering
    \includegraphics[width=1.333\columnwidth]{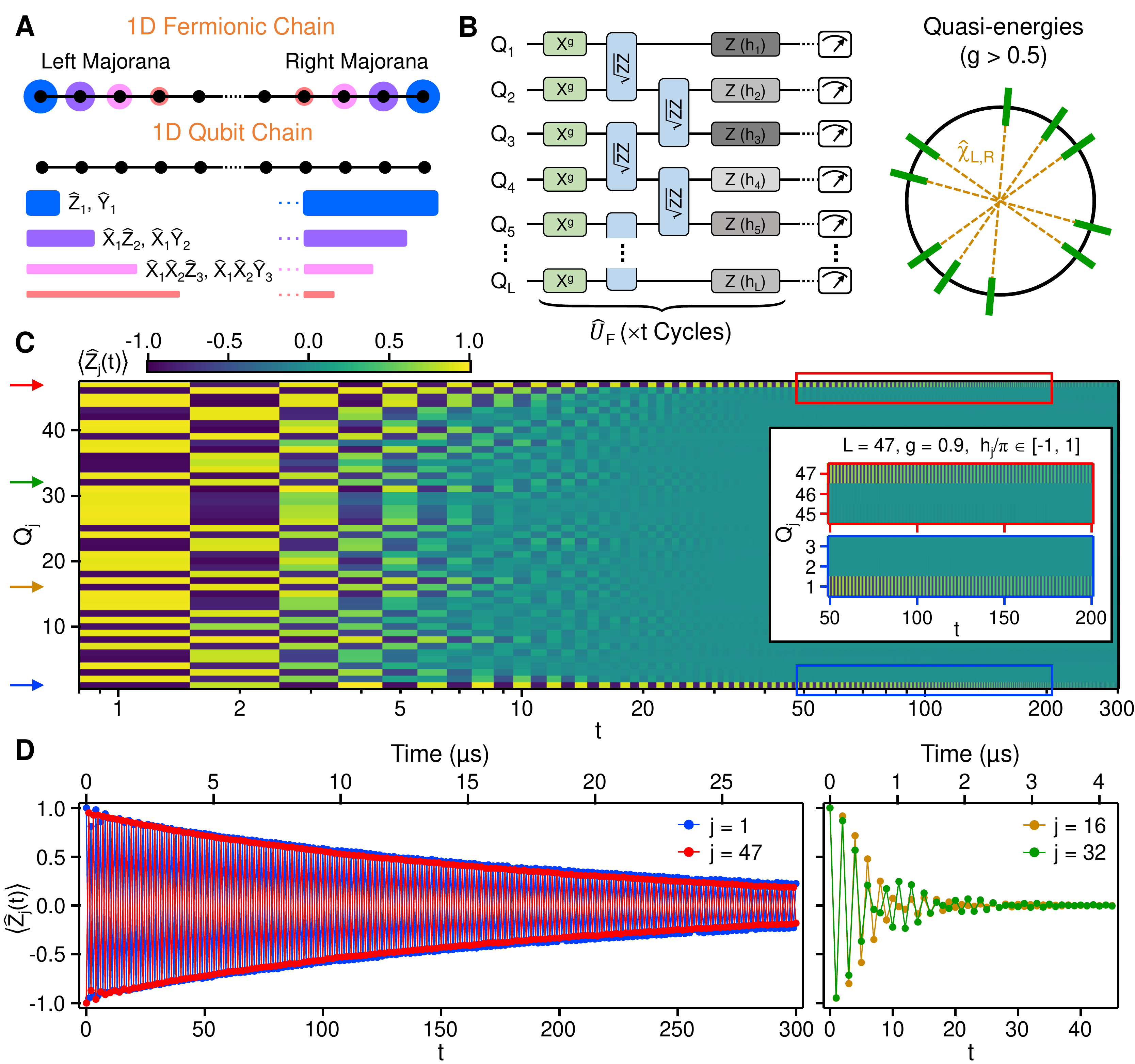} 
    \caption{Observation of long-lived edge modes in a kicked Ising model. (A) Schematic illustration of the Jordan-Wigner transformation between a 1D fermionic Kitaev chain and a qubit chain. In the fermionic (qubit) chain, the sizes (widths) of the colored spheres (bars) denote the relative weights of the edge modes in the Majorana fermion (Pauli) basis. The right edge mode in the Pauli basis is dominated by long Pauli operators spanning the entire chain. (B) Left: Quantum circuit implementation of a kicked Ising model. An identical unitary $\hat{U}_\text{F}$ is repeated a total of $t$ times. Right: Eigenstates of $\hat{U}_\text{F}$ ($g > 0.5$), shown on a unit circle according to their quasienergies. (C) $\braket{\hat{Z}_j (t)}$ as a function of $t$ and qubit location $Q_j$. The initial state is a random product state, $\ket{0101001...}$. Inset shows $\braket{\hat{Z}_j (t)}$ for the three leftmost and rightmost qubits, between $t = 50$ and $t = 200$. (D) $\braket{\hat{Z}_j (t)}$ for the two edge qubits $j = 1, 47$ (left panel) and two qubits within the bulk $j = 16, 32$ (right panel). Top axis for each plot indicates real time, calculated based on the time needed to execute $\hat{U}_\text{F}$ (93 ns). Locations for the qubits shown in this panel are also indicated by colored arrows in panel C.}
    \label{fig:1}
\end{figure*}

The advent of high-fidelity quantum processors and simulators suggests an alternative approach to examine the realistic extent of protection for a given symmetry \cite{Blatt_NatPhys_2012,Gross_Science_2017,Carusotto_NatPhys_2020}. Here we use Jordan-Wigner transformation (JWT) to map the Kitaev model to a transverse Ising spin model \cite{LIEB1961407}, which is more compatible with a chain of qubits~\cite{LevitovMooij_arxiv2021,Nori2014,Guo_Sci_Adv_2018,Dykman_PRA_2019}. The JWT also maps each MEM, commonly represented by a sum of local Majorana operators in the fermionic chain, to a sum of Pauli spin operators that can be individually characterized on a quantum processor. Given the non-local nature of the JWT, the MEMs in the Pauli basis are prone to local symmetry-breaking noise even within a closed system, setting them distinct from MEMs in fermionic systems. Despite this disadvantage, we find that the interplay between $\mathbb{Z}_2$ parity symmetry and a prethermalization mechanism endows the MEMs with a strong resilience toward both closed-system thermalization and open-system perturbations such as low-frequency noise. Furthermore, we discover a method for accurately reconstructing the Pauli expansion of MEMs in the presence of decoherence, which may be extended to study other integrals of motion in many-body quantum systems.

\begin{figure*}[t!]
    \centering
    \includegraphics[width=2\columnwidth]{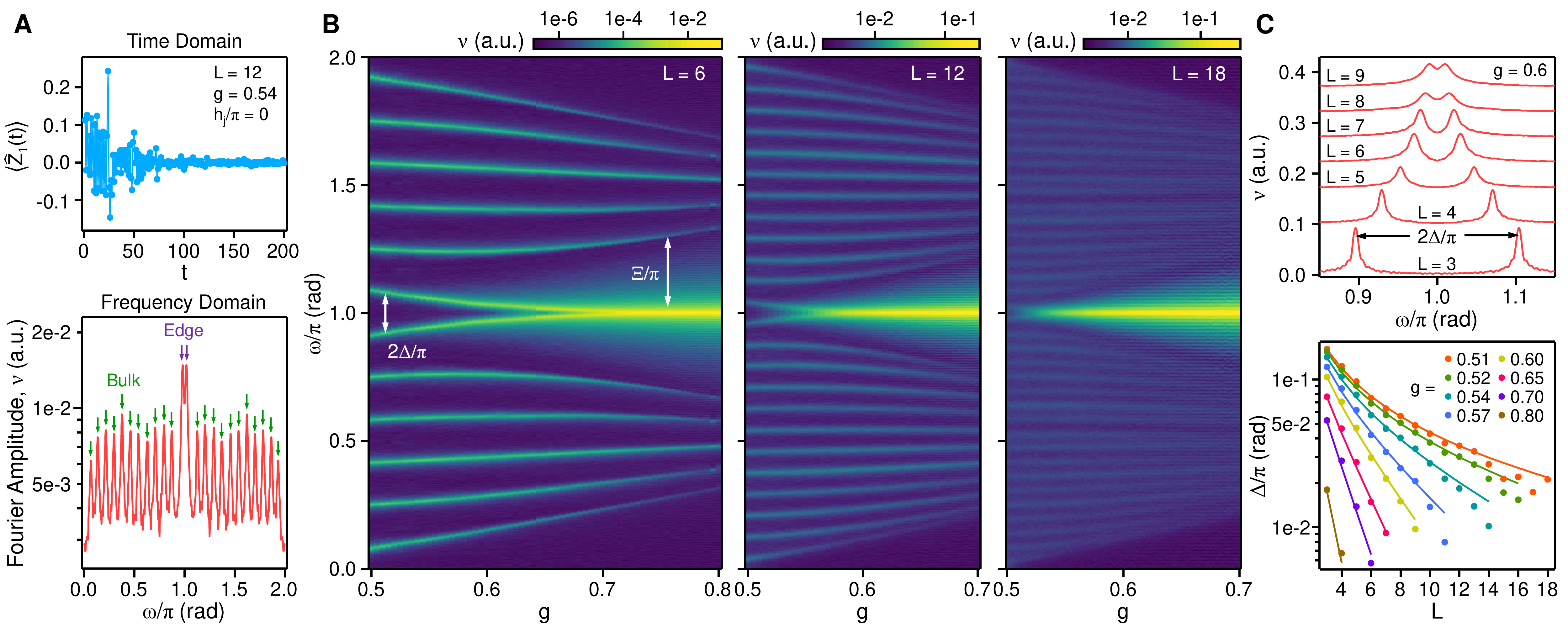} 
    \caption{Quasienergy spectroscopy. (A) Top: $\braket{\hat{Z}_1 (t)}$ measured in the integrable limit $h_j = 0$ and system size $L = 12$. Bottom: Frequency-dependent amplitude $\nu (\omega)$ of the Fourier transform of $\braket{\hat{Z}_1 (t)}$. The arrows indicate the single-particle quasienergy peaks for the bulk and edge fermionic modes. (B) $\nu$ as a function of both frequency $\omega/\pi$ and $g$, measured for three different values of $L$. $h_j = 0$ in all cases. To obtain the spectra, $\braket{\hat{Z}_1 (t)}$ is measured up to $t = $ 300, 200 and 150 cycles for $L = 6$, 12 and 18, respectively. (C) Top: $\nu (\omega)$ for different $L$, showing the quasienergy peaks of the hybridized MEMs with a splitting $2 \Delta / \pi$. Data are offset for clarity. Bottom: $\Delta / \pi$ measured as a function of $L$ at different values of $g$. Solid lines represent exact numerical results from diagonalizing $\hat{U}_\text{F}$ in the fermionic basis \cite{SM}. Random product states are used as  initial states in all measurements.}
    \label{fig:2}
\end{figure*}

The experiment is conducted on an open-ended chain of $L = 47$ superconducting qubits (see Supplementary Materials (SM) for device details \cite{SM}). The qubit chain is periodically driven by a quantum circuit corresponding to a kicked Ising model (Fig.~\ref{fig:1}B), with the following unitary applied in each cycle:
\begin{equation}
    \hat{U}_\text{F}=\,\scaleto{e}{6pt}^{\scaleto{-\cfrac{i}{2}\displaystyle\sum_{j=1}^L h_j \hat{Z}_j}{28pt}}\, \scaleto{e}{6pt}^{\scaleto{-\cfrac{i\pi J}{2} \displaystyle\sum_{j=1}^{L-1} \hat{Z}_j \hat{Z}_{j+1}}{28pt}}\, \scaleto{e}{6pt}^{\scaleto{-\cfrac{i\pi g}{2} \displaystyle\sum_{j=1}^L \hat{X}_j}{28pt}}, 
\end{equation}
\noindent
where $\hat{X}_j$, $\hat{Y}_j$ and $\hat{Z}_j$ denote Pauli operators acting on a given qubit $Q_j$. Compared to a digitized implementation of the transverse Ising Hamiltonian (see Fig.~S11 of the SM for experimental data of this approach \cite{SM}), the periodic (a.k.a. Floquet) evolution here generates faster dynamics in real time and is advantageous due to the finite coherence times of the qubits. Under such a drive, the Hilbert space of the system may be described by the eigenstates of $\hat{U}_\text{F}$ whose eigenvalues lie on a unit circle, as illustrated in the right panel of Fig.~\ref{fig:1}B. 

Given there is no ground state in a Floquet system, the two-fold degeneracy of the ground state in the Kitaev model becomes instead a pairing of eigenstates across the entire spectrum. In our work, we fix $J = 1/2$ wherein the Floquet system, in the integrable limit $h_j = 0$, has $\mathbb{Z}_2$ spin-flip symmetry and exhibits two phases with distinct spectral pairings: At $g > 0.5$ which is the focus of the main text, the quasienergy levels have a $\pi$-pairing (right panel of Fig.~\ref{fig:1}B): Every many-body eigenstate of the $\hat{U}_\text{F}$ with quasienergy $\theta$ has a ``partner'' state with quasienergy $\theta+\pi$ \cite{Khemani_PRL_2016}. The transition between any paired eigenstates is enabled by an application of the so-called $\pi$-MEMs, $\hat{\chi}_\text{L}$ and $\hat{\chi}_\text{R}$ \cite{DuttaPRB13,Mitra19}. The $\pi$-MEMs anti-commute with $\hat{U}_\text{F}$ in the large $L$ limit:
\begin{equation}
\hat{\chi}_\text{L} \hat{U}_\text{F} = - \hat{U}_\text{F} \hat{\chi}_\text{L}\,, \,\,\text{  } \hat{\chi}_\text{R} \hat{U}_\text{F} = - \hat{U}_\text{F} \hat{\chi}_\text{R},
\label{eqn:commutation}
\end{equation}
At $g < 0.5$, the eigenspectrum of $\hat{U}_\text{F}$ has a double degeneracy: each eigenstate has a partner state with the same quasienergy. Here the transition between paired eigenstates is described by two so-called $0$-MEMs which commute with  $\hat{U}_\text{F}$. Experimental data for this regime, which is analogous to the ferromagnetic phase of the transverse Ising model, are shown in Fig.~S10 of the SM \cite{SM}. Lastly, at a critical point $g = 0.5$, the eigenstates are distributed uniformly on the unit circle with a gap of $\pi / L$, which vanishes in the limit $L = \infty$.

In the presence of finite local fields $h_j \neq 0$, $\hat{U}_\text{F}$ is no longer integrable and the $\mathbb{Z}_2$-symmetry is also broken. We begin by searching for signatures of stable edge modes in this regime and focusing on the $\hat{Z}$ operators which, in the JWT, have large overlap with MEMs on the edge (Fig.~\ref{fig:1}A). Figure~\ref{fig:1}C shows experimental measurements of $\braket{\hat{Z}_j (t)}$ for all qubits in the chain, where we have chosen $h_j / \pi$ from a random uniform distribution $[-1, 1]$ to maximize the effect of integrability-breaking. We observe a stark contrast in behavior of the edge qubits, $Q_{1}$ and $Q_{47}$, and qubits within the chain, $Q_{2}$ to $Q_{46}$. Whereas $\braket{\hat{Z}_j (t)}$ decays rapidly to 0 after $\sim$20 cycles ($\sim$2 $\mu$s) for any qubit in the bulk, $\braket{\hat{Z}_j (t)}$ decays much more slowly for the edge qubits. In addition, $\braket{\hat{Z}_j (t)}$ for each edge qubit shows a subharmonic oscillation at a period twice that of the drive $\hat{U}_\text{F}$, since each application of $\hat{U}_\text{F}$ changes the sign of $\hat{\chi}_\text{L, R}$ due to their anti-commutation (Eqn.~\ref{eqn:commutation}). 

The bulk-edge difference is further illustrated in Fig.~\ref{fig:1}D, where data for four qubits are shown. The lifetimes of the edge modes, which include contributions from both external decoherence effects and internal non-integrable dynamics, are extracted by fitting the envelope of $\braket{\hat{Z}_1 (t)}$ ($\braket{\hat{Z}_{47} (t)}$) to an exponential (Fig.~S8 of the SM \cite{SM}) and found to be 19.5 $\mu$s (17.2 $\mu$s) for $Q_1$ ($Q_{47}$). These values are close to the typical single-qubit relaxation time $T_1 = 22.2$ $\mu$s on the device -- a preliminary indication that the MEMs are resilient toward integrability- and symmetry-breaking fields as well as dephasing effects such as low-frequency noise.

Recent theoretical works have suggested that the resilience of the edge modes toward non-integrable dynamics is a result of prethermalization \cite{Fendley2016, PrethermalRigorous,MoriPRL16_RigorousBoundHeating,ElsePrethermalTimeCrystalPRX,FendleyPRXPreth}. Unlike thermalizing systems which monotonically decay to ergodic states over time, a prethermal system relaxes first to a meta-stable state before decaying to ergodic states. A common mechanism for prethermalization is the existence of spectral gaps which make relaxation processes driven by integrability-breaking perturbations off-resonant, thereby preventing energy absorption. To experimentally establish prethermalization in our system, we characterize the excitation spectrum in the integrable limit, $h_j = 0$. Here the many-body spectrum of $\hat{U}_F$ may be constructed from a total of $2L$ energy quanta, corresponding to the quasienergies of non-interacting Bogoliubov fermionic quasiparticles in the fermionic representation of $\hat{U}_F$. These quasienergies can be obtained via a Fourier analysis of time-domain signals \cite{roushan2017} (see Sections III and V of the SM \cite{SM}). Figure~\ref{fig:2}A shows measurements of $\braket{\hat{Z}_1 (t)} (h_j = 0)$ for a short chain $L = 12$. The time evolution for $\braket{\hat{Z}_1 (t)}$ is now seemingly featureless, which results from interference between different eigenmodes of $\hat{U}_F$. To obtain the quasienergies, a Fourier transform of the time-domain data is then performed, with the results also shown in Fig.~\ref{fig:2}A. The Fourier spectrum $\nu (\omega)$ reveals a total of $2L$ distinct peaks at values of $\omega$ corresponding to the quasienergies of the $2L$ non-interacting fermionic modes in the system.

\begin{figure*}[t!]
    \centering
    \includegraphics[width=2\columnwidth]{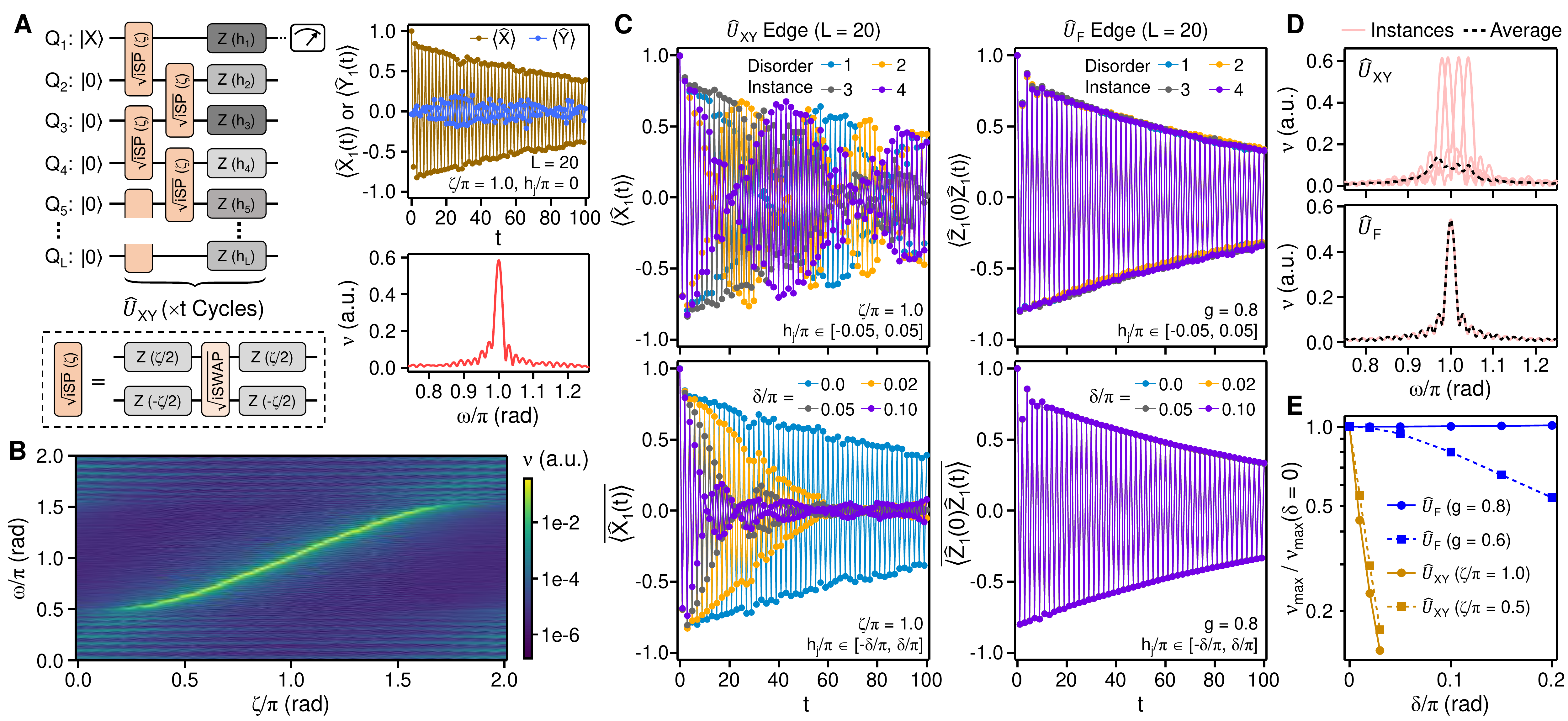} 
    \caption{Low-frequency noise resilience of MEMs and comparison with unprotected edge modes. (A) Left: Quantum circuit corresponding to the XY model, where an identical cycle unitary $\hat{U}_\text{XY}$ is applied $t$ times. Right: Top panel shows $\braket{\hat{X}_1 (t)}$ and $\braket{\hat{Y}_1 (t)}$ measured at $Q_1$, with the control parameter $\zeta / \pi = 1.0$ and no disorder $h_j / \pi = 0$. Lower panel shows the Fourier spectrum $\nu (\omega)$ of $\braket{\hat{X}_1 (t)} + i \braket{\hat{Y}_1 (t)}$. (B) $\nu(\omega)$ as a function of $\omega/\pi$ and $\zeta$ for the $\hat{U}_\text{XY}$ model, where $\braket{\hat{X}_1 (t)}$ and $\braket{\hat{Y}_1 (t)}$ are measured up to $t = 100$. (C) Top panels: $\braket{\hat{X}_1 (t)}$ ($\braket{\hat{Z}_1 (0) \hat{Z}_1 (t)}$) for the $\hat{U}_\text{XY}$ ($\hat{U}_\text{F}$) edge modes, measured for 4 different disorder realizations with $h_j / \pi \in [-0.05, 0.05]$. Bottom panels: Disorder-averaged $\overline{\braket{\hat{X}_1 (t)}}$ $\left( \overline{\braket{\hat{Z}_1 (0) \hat{Z}_1 (t)}} \right)$ for the $\hat{U}_\text{XY}$ ($\hat{U}_\text{F}$) edge mode, shown over 4 different disorder strengths $\delta$. 80 disorder instances $h_j / \pi \in [-\delta / \pi, \delta / \pi]$ are used for averaging in each case, and the initial states are additionally randomized between instances for $\hat{U}_\text{F}$. (D) Red lines: Fourier spectra $\nu (\omega)$ obtained from the disorder instances in the upper panels of C. Black lines: $\nu (\omega)$ for the disorder-averaged observables ($\delta = 0.05$) in the lower panels of C. (E) Maximum Fourier amplitude $\nu_\text{max}$ as a function of $\delta$. Data are normalized by $\nu_\text{max}$ at $\delta = 0$.}
    \label{fig:3}
\end{figure*}

The two dominant peaks close to $\omega = \pi$ in the spectrum of Fig.~\ref{fig:2}A are associated with the MEMs, which are split in quasi-energy due to their hybridization in this short chain. To confirm this interpretation, we change the localization length $\xi$ of the MEMs by tuning $g$ and measure the spectra over three different system sizes. The results, shown in Fig.~\ref{fig:2}B, reveal two important features: First, we observe that the quasienergy splitting $2 \Delta$ of the two MEMs decreases as $g$ increases. This is due to a reduced $\xi$ that leads to weaker hybridization between $\hat{\chi}_\text{L}$ and $\hat{\chi}_\text{R}$. Second, we observe a finite quasienergy gap $\Xi$ between the MEMs and the other bulk fermionic modes, which also increases at larger $g$. This quasienergy gap, which crucially remains open as $L$ increases, suppresses transitions between bulk and edge states and is the key to protecting the MEMs against integrability-breaking fields. Further discussion of this prethermalization mechanism is presented in Section IV of the SM \cite{SM}.

While the bulk gap protects the MEMs from internal thermalization, the finite quasi-energy difference $2\Delta$ between the two MEMs is sensitive to disorder fluctuations. Such a sensitivity may lead to dephasing of the MEMs through low-frequency noise \cite{Knapp_PRB_2018}, as shown by Fig.~S5 in the SM \cite{SM}. This effect may be suppressed by reducing the hybridization between $\hat{\chi}_\text{L}$ and $\hat{\chi}_\text{R}$, which is achievable through increasing either $g$ (Fig.~\ref{fig:2}B) or $L$. The dependence of $\Delta$ on $L$ is mapped out in detail by the experimental measurements shown in Fig.~\ref{fig:2}C. We observe that for $g > 0.6$, $\Delta$ is exponentially suppressed by larger $L$, in agreement with theory \cite{SM}. For $g < 0.6$, the suppression is no longer exponential due to proximity to the phase transition point $g = 0.5$ where the bulk gap closes. We also find excellent agreement between exact numerical results and experimental measurements even for $\Delta / \pi \approx 0.01$, which is a result of accurate gate calibrations described in the SM \cite{SM}.

\begin{figure*}[t!]
\centering
\includegraphics[width=1.333\columnwidth]{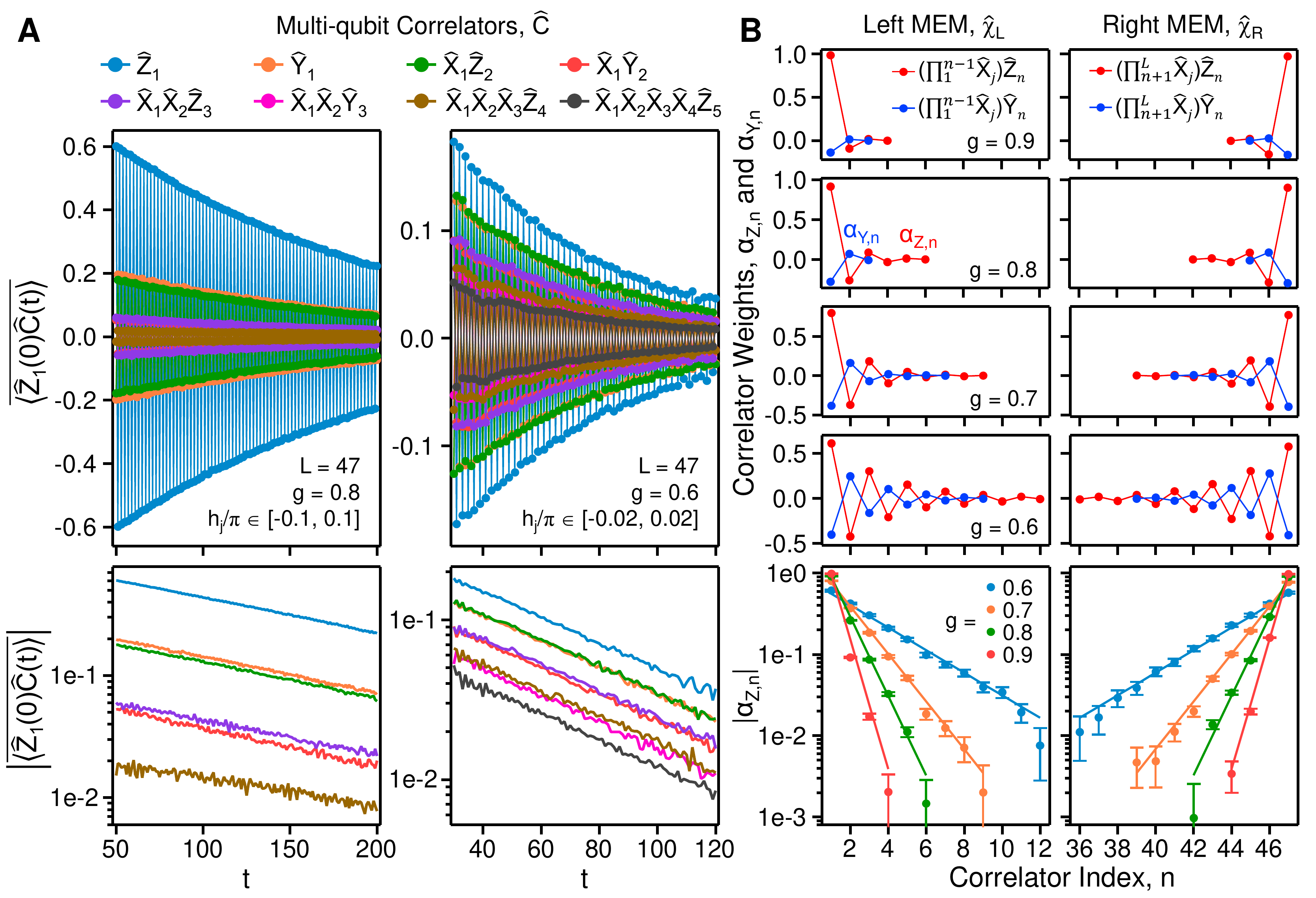} 
\caption{Reconstructing the Pauli operator expansion of MEMs. (A) Upper panels: Correlators $\overline{\braket{\hat{Z}_1 (0) \hat{C}(t)}}$ for $g = 0.8$ and $g = 0.6$, with the compositions of $\hat{C}(t)$ shown in the legend. Here the $g = 0.8$ ($g = 0.6$) data are averaged over 10 (12) disorder realizations and initial random product states. Lower panels show the absolute values of the correlators, $\left| \overline{\braket{\hat{Z}_1 (0) \hat{C}(t)}} \right|$. (B) Top 8 panels show experimentally reconstructed Pauli operator expansion of the MEMs $\hat{\chi}_\text{L, R}$. $\alpha_{\text{Z},n}$ and $\alpha_{\text{Y},n}$ correspond to the coefficients of the Pauli operators shown in the legends. Bottom panels show experimental values of $|\alpha_{\text{Z},n}|$ (points) and theoretical predictions (solid lines). Error bars correspond to statistical uncertainty due to single-shot measurements (see Methods in SM \cite{SM}).}
\label{fig:4}
\end{figure*}

We now perform a systematic study on the low-frequency noise resilience of the MEMs for a moderately long chain, $L = 20$. To examine the role of symmetry in noise protection, we have also experimentally realized edge modes in a different periodic circuit with a cycle unitary $\hat{U}_\text{XY}$ that does not possess $\mathbb{Z}_2$-symmetry. As illustrated in Fig.~\ref{fig:3}A, $\hat{U}_\text{XY}$ consists of two layers of two-qubit gates applied between all nearest-neighbor qubits, $\sqrt{\text{iSP} (\zeta)} = e^{-i\frac{\zeta}{4} \left( \hat{Z}_j - \hat{Z}_{j+1} \right)} e^{-i\frac{\pi}{4} \left( \hat{\sigma}^+_j \hat{\sigma}^-_{j+1} + \hat{\sigma}^-_j \hat{\sigma}^+_{j+1} \right)} e^{-i\frac{\zeta}{4} \left( \hat{Z}_j - \hat{Z}_{j+1} \right)}$, where $\hat{\sigma}^{+,-}$ denotes Pauli raising and lowering operators. In the single-excitation subspace, $\hat{U}_\text{XY}$ has $L$ eigenmodes including two localized edge modes (see Appendix E of Ref.~\cite{Neill_Nature_2021}) for control parameter $\zeta / \pi \in [0.25, 1.75]$. The leading order terms in the Pauli operator expansion of the edge modes are $\hat{\sigma}^+_{1}$ and $\hat{\sigma}^+_\text{L}$, respectively.

To probe one of the edge modes for $\hat{U}_\text{XY}$, we prepare the system in a superposition state $\frac{1}{\sqrt{2}} \left( \ket{0000...} + \ket{1000...}\right)$ to maximize the initial value $\braket{\hat{\sigma}^+_{1} (t = 0)} = 1$. We then apply $\hat{U}_\text{XY}$ $t$ times before measuring the time-dependent observable $\braket{\hat{\sigma}^+_{1} (t)} = \braket{\hat{X}_1 (t)} + i\braket{\hat{Y}_1 (t)}$, which precesses at a frequency corresponding to the quasienergy of the edge mode. Example experiment data and the corresponding Fourier spectrum $\nu (\omega)$ for $\zeta / \pi = 1.0$ are both shown in Fig.~\ref{fig:3}A, demonstrating a slowly decaying subharmonic response and a quasienergy peak at $\omega = \pi$ that are similar to the MEMs of the $\hat{U}_\text{F}$ model. Figure~\ref{fig:3}B shows experimentally measured $\nu$ as a function of both $\zeta$ and $\omega$. At $0.25 \lesssim \zeta / \pi \lesssim 1.75$, we observe a dominant quasienergy peak which corresponds to an edge mode and is separated from the $L - 2$ bulk modes, visible as smaller peaks outside the range $0.5 < \omega_\pi < 1.5$, by spectral gaps akin to the bulk gap $\Xi$ of the $U_\text{F}$ model. Despite these apparent similarities, a crucial distinction exists between the two models: the quasienergy of the $\hat{U}_\text{XY}$ edge mode is first-order sensitive to $\zeta$ at all values of $\zeta$ whereas the quasienergy of the $\hat{U}_\text{F}$ edge mode asymptotically approaches $\pi$ as $g$ increases. This distinction is due to the lack of $\mathbb{Z}_2$-symmetry in $\hat{U}_\text{XY}$ and leads to dramatically different robustness of the two models toward low-frequency noise in $h_j$, which we explore next.

The upper panels of Fig.~\ref{fig:3}C show $\braket{\hat{X}_1 (t)}$ ($\braket{\hat{Z}_1 (0) \hat{Z}_1 (t)}$) of the $\hat{U}_\text{XY}$ ($\hat{U}_\text{F}$) model, measured for four different realizations of $h_j / \pi$ which are uniformly chosen from $[-\delta, \delta]$. Here the autocorrelator $\braket{\hat{Z}_1 (0) \hat{Z}_1 (t)}$ differs from $\braket{\hat{Z}_1 (t)}$ only by a random $\pm$ sign given by the initial state of $Q_1$, and $\delta = 0.05$ is a disorder strength chosen to be comparable to the low-frequency fluctuation of the quantum device. We observe that $\braket{\hat{X}_1 (t)}$ exhibits beating patterns that depend sensitively on the disorder realization. This is a result of the first-order sensitivity toward control parameters demonstrated in Fig.~\ref{fig:3}B. On the other hand, $\braket{\hat{Z}_1 (0) \hat{Z}_1 (t)}$ is virtually unchanged between different disorder realizations. The impact of low-frequency noise on each edge mode realization is then emulated by averaging the corresponding observable over an ensemble of disorder realizations, which mimics the process of dephasing. The disorder-averaged $\overline{\braket{\hat{X}_1 (t)}}$ in the $\hat{U}_\text{XY}$ model, shown in the lower panel of Figure~\ref{fig:3}C, decays significantly faster as the disorder strength $\delta$ increases. On the other hand, $\overline{\braket{\hat{Z}_1 (0) \hat{Z}_1 (t)}}$ in the $\hat{U}_\text{F}$ model remains unchanged over $\delta$. 

The sensitivity of the two edge mode realizations toward low-frequency noise is further elucidated by inspecting the Fourier spectrum $\nu (\omega)$ of each disorder realization, shown in Fig.~\ref{fig:3}D. Here we observe that the quasienergy peak for the $\hat{U}_\text{XY}$ edge mode is different for each disorder realization, resulting in a broadened spectrum with a lower peak height upon averaging. On the other hand, the quasienergy peak for $\hat{U}_\text{F}$ remains stable at $\omega = \pi$, irrespective of disorder realizations. Lastly, we measure the disorder-averaged quasienergy peak height, $\nu_\text{max} = \text{Max} [\nu (\omega)]$, and show the results in Fig.~\ref{fig:3}E. For the $\hat{U}_\text{XY}$ model, we observe that $\nu_\text{max}$ decays exponentially as a function of $\delta$ irrespective of $\zeta$. For the $\hat{U}_\text{F}$ model, $\nu_\text{max}$ is completely insensitive to $\delta$ for sufficiently localized MEMs ($g = 0.8$) and remains insensitive for small $\delta < 0.05$ even in the more delocalized regime $g = 0.6$. These results highlight the critical role of symmetry in stabilizing the quasienergies of MEMs and protecting their lifetimes against low-frequency noise.

Finally, using the full $L = 47$ qubit chain, we demonstrate an error-mitigation strategy for accurately reconstructing the Pauli operator expansion of $\hat{\chi}_\text{L, R}$ in the presence of noise. Figure~\ref{fig:4}A shows the late-time evolution of eight multi-qubit Pauli operators $\hat{C}$ entering the JWT of $\hat{\chi}_\text{L, R}$, experimentally obtained by rotating each qubit into the appropriate basis followed by multi-qubit readout. We observe that each $\overline{\hat{Z}_1 (0) \hat{C} (t)}$ exhibits a similar subharmonic response with an amplitude that decreases when $\hat{C}$ incorporates more qubits and has less overlap with $\hat{\chi}_\text{L, R}$ (Fig.~\ref{fig:1}A). Operators ending with $\hat{Y}$ also show smaller amplitudes than those ending with $\hat{Z}$ since they have no overlap with $\hat{\chi}_\text{L, R}$ in the time-independent transverse Ising model and only arise as corrections to the JWT of $\hat{\chi}_\text{L, R}$ due to the time-dependent, periodic dynamics. Strikingly, as shown also in Fig.~\ref{fig:4}A, the absolute values (i.e. magnitudes) of these operators, $\left| \overline{\hat{Z}_1 (0) \hat{C} (t)} \right|$, exhibit nearly identical decay rates despite their different lengths and compositions. 

The observation in Fig.~\ref{fig:4}A is contrary to naive expectations, wherein the decay rate of a quantum operator is expected to scale with the number of qubits it incorporates. The result may be qualitatively understood by the fact that $\hat{\chi}_\text{L}$ and $\hat{\chi}_\text{R}$ anti-commute with $\hat{U}_\text{F}$ (Eqn.~\ref{eqn:commutation}) and are conserved under the periodic dynamics. Even though external decoherence and integrability-breaking fields violate this commutation, $\hat{\chi}_\text{L, R}$ remains a slowly decaying mode (see Section VII of the SM \cite{SM}). As a result, any multi-qubit operator having a finite overlap with $\hat{\chi}_\text{L, R}$ will exhibit a slow-decaying expectation value in its late-time dynamics, with an amplitude proportional to the overlap. 

The uniform decay rates of $\overline{\hat{Z}_1 (0) \hat{C} (t)}$ inform an experimental strategy for reconstructing the expansion of the MEMs in the Pauli operator basis:
\begin{equation}
\scaleto{\hat{\chi}_\text{L, R} = \sum_{n = 1}^L \left[ \alpha_{\text{Z}, n} \left( \prod_{j = M}^N \hat{X}_j \right) \hat{Z}_n +  \alpha_{\text{Y}, n} \left( \prod_{j = M}^N \hat{X}_j \right) \hat{Y}_n \right]}{32pt}, 
\end{equation}
where the products over $j$ have limits $N = n - 1$ and $M = 1$ for $\hat{\chi}_\text{L}$, and $N = L$ and $M = n + 1$ for $\hat{\chi}_\text{R}$ \cite{note}. The coefficients $\alpha_{\text{Z}, n}$ and $\alpha_{\text{Y}, n}$ normalize to unity for $L = \infty$: $\sum_n | \alpha_{\text{Z}, n}| ^ 2 + |\alpha_{\text{Y}, n}| ^ 2 = 1$. To estimate their values, we measure different $\overline{\hat{Z}_1 (0) \hat{C} (t)}$ at 10 late-time cycles. The average value of each operator and the normalization condition allow us to determine the ideal values of $\alpha_{\text{Z}, n}$ and $\alpha_{\text{Y}, n}$ (see Methods and Section VI of the SM for details \cite{SM}). 

The experimentally measured coefficients, shown in Fig.~\ref{fig:4}B for four values of $g$, both oscillate in sign and decay exponentially as $n$ moves away from the edge. The decay rate is also observed to decrease as $g$ approaches the critical value $g = 0.5$. This is due to that fact that the decay constant for the coefficients is the localization length $\xi$ of the MEMs, which diverges at $g = 0.5$ (see Section III of the SM \cite{SM}). A comparison between theoretical and experimental values of $|\alpha_{\text{Z}, n}|$ is shown in the bottom panels of Fig.~\ref{fig:4}B, where good agreement is found over a span of nearly three orders of magnitude.

In conclusion, we simulate MEMs using a system of driven transmon qubits and comprehensively study their symmetry protection against noise in their solid-state environment. We find the degree of protection sensitively depends on the physical characteristic of the noise and generally does not extend to noise that breaks the underlying symmetry, such as $T_1$ decay of the transmon qubits. Interestingly, we also find that owing to a prethermalization mechanism, the MEMs in our system are also protected against certain noise that seemingly violates $\mathbb{Z}_2$ symmetry, e.g. local $\hat{Z}$ fluctuations. These results highlight the complex interplay between physical noise and protection, and indicate the crucial importance of testing symmetry against open-system dynamics in any experimental platform. Furthermore, we find that even in the presence of decoherence, the Pauli expansion of conserved quantities such as MEM can be accurately determined by measuring and re-normalizing late-time expectation values of Pauli operators. This error-mitigation strategy may be applied to study integrals of motion in physical models more difficult to compute classically. Preliminary results on non-integrable dynamics are shown in Fig.~S7 of the SM \cite{SM}).

{\bf Acknowledgements--- } We have benefited from discussions with M.~H.~Devoret, L.~G.~Dias, I.~K.~Drozdov, P.~Ghaemi, and A.~Rahmani. D.~Bacon is a CIFAR Associate Fellow in the Quantum Information Science Program.

{\bf Author contributions---} D.~A.~Abanin and V.~Smelyanskiy conceived the project. X.~Mi, D.~A.~Abanin and V.~Smelyanskiy designed the experiment. X.~Mi executed the experiment. P.~Roushan, X.~Mi, D.~A.~Abanin, M.~Sonner and M.~Niu performed analysis of the experimental results. K.~W.~Lee and B.~Foxen contributed to measurements in the Supplementary Materials. X.~Mi, P.~Roushan and D.~A.~Abanin wrote the manuscript. V.~Smelyanskiy and P.~Roushan led and coordinated the project. Infrastructure support was provided by Google Quantum AI. All authors contributed to revising the manuscript and the Supplementary Materials.

{\bf Data availability--- }The experimental data contained in the main text and Supplementary Materials will be included with final publication of the manuscript.

{\bf Code availability--- }The Python simulation code used in theoretical analysis will be included with the final publication of the manuscript.

{\bf Competing interests--- }The authors declare no competing interests.

\twocolumngrid
\bibliography{mbl.bib}

\begin{thebibliography}{10}
\bibitem{note}
In Eqn~3, product terms $\prod_{j = M}^N \hat{X}_j$ with $N < M$ should be
  treated as identity $\hat{I}$. Additionally, the JWT of $\hat{\chi}_\text{R}$
  should, in principle, have higher weights in longer Pauli strings, with the
  dominant string being $\left( \prod_{j=1}^{L-1} \hat{X}_j \right) \hat{Y}_L$.
  This is equivalent to $\hat{Z}_L$ up to the parity transformation
  $\prod_{j=1}^{L} \hat{X}_j$ which commutes with $\hat{U}_\text{F}$. In
  Eqn.~3, $\hat{\chi}_\text{R}$ is expressed in this alternative form and has a
  leading term $\alpha_{\text{Z}, L} \hat{Z}_L$. Such a basis requires
  measurements of shorter Pauli strings.
\end{thebibliography}
\bibliographystyle{science.bst}

\onecolumngrid

\newpage
\textbf{Materials and Methods}

Quantum processor details are described in Section I of the Supplementary Materials. The $\sqrt{ZZ}$ gates used in this work are implemented with a combination of two-qubit CZ gates and local $Z$ gates.

For reconstructing the operator expansion in Fig.~4B of the main text, we measure different correlators at ten fixed cycles and average the results (while accounting for the alternating $\pm$ signs of the correlators). Higher order correlators that are too small to resolve in experiment (i.e. when it is $\lessapprox 10^{-3}$) are not measured. The cycles used are $t = 170$ to 180 for $g = 0.9$, $t = 140$ to 150 for $g = 0.8$, $t = 120$ to 130 for $g = 0.7$ and $t = 90$ to 100 for $g = 0.6$. The average values of the measured correlators are then re-scaled by their norm, $A_\text{norm} = \sqrt{\sum_{\{ \hat{C} \}} \left| \overline{\braket{\hat{Z}_1 (0) \hat{C} (t)}} \right|^2}$. Here the overline denotes further averaging over 10 random instances of initial states and disorders in $h_j / \pi \in [ -\delta / \pi, \delta / \pi]$, where $\delta/\pi$ is chosen to be 0.1 for $g = 0.9$ and $g = 0.8$, and 0.02 for $g = 0.7$ and $g = 0.6$. The error bars in Fig.~4B of the main text are estimated based on the number of measurement shots, $N_\text{shots}$, which gives a statistical uncertainty of $1/\sqrt{N_\text{shots}}$. This uncertainty is further amplified by the normalization procedure to a final value of $1/(A_\text{norm} \sqrt{N_\text{shots}})$.

\tableofcontents

\section{Quantum processor details}

\subsection{Device details, coherence times and gate fidelities}

\begin{figure}[t!]
  \centering
  \includegraphics[width=1\columnwidth]{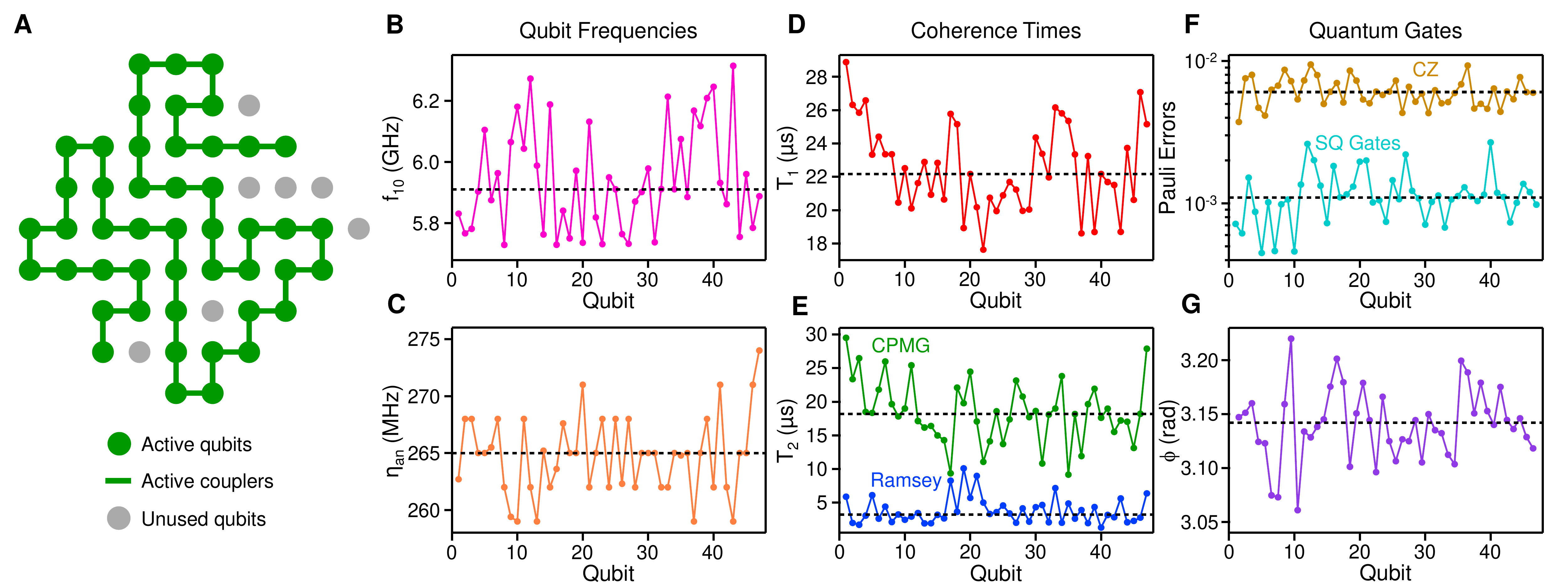} 
 \caption{Device details, coherence times and gate fidelities. (A) Layout of a 54-qubit quantum processor on which the experiment is conducted. The active qubits and couplers used to form the 47-qubit chain are indicated by the green dots and lines. Inactive qubits are indicated by gray dots. (B) Qubit frequency $f_{10}$ as a function of qubit location on the 47-qubit chain. (C) Qubit anharmonicity $\eta_\text{an}$ as a function of qubit location on the 47-qubit chain. (D) Single-qubit $T_1$ vs qubit location. (E) Single-qubit $T_2$ vs qubit location, obtained by Ramsey interferometry ($T_2^*$) and Carr-Purcell-Meiboom-Gill (CPMG) dynamical decoupling ($T_2^\text{CPMG}$). (F) Pauli error rates for both single-qubit and CZ gates. Error rates for single-qubit gates are obtained through randomized benchmarking (RB) with Clifford gates. Error rates for CZ are obtained from cross-entropy benchmarking (XEB) with random circuits comprising cycles of interleaved CZ and single-qubit rotations, similar to previous experiments \cite{Arute2019,Mi_OTOC_2021,DTC_Nature_2022}. Single-qubit error rates are then subtracted from the resulting cycle errors to arrive at the CZ error rates. (G) Conditional phase $\phi$ for each CZ gate, obtained from Floquet calibration \cite{DTC_Nature_2022}. All measurements in panels A through D are conducted in parallel across the 47-qubit chain. Dashed lines indicate the median values of different metrics.}
 \label{fig:s1}
\end{figure}

The quantum processor used in our experiments consists of a 2D grid of 54 superconducting transmon qubits that have both tunable frequencies and tunable interqubit couplings, similar in design to the Sycamore processor used in Ref.~\cite{Arute2019}. The exact layout of the quantum processor is shown in Fig.~\ref{fig:s1}A. A 1D chain of 47 qubits are chosen from the 2D grid and used to conduct the experiments described in the main text and the rest of Supplementary Materials (SM). The frequencies of the transmon qubits along the chain are shown in Fig.~\ref{fig:s1}B, where a median value (indicated by the dashed line) of 5.91 GHz is seen. The qubit anharmonicity ($\eta_\text{an}$), defined as the difference between the $\ket{0} \rightarrow \ket{1}$ transition frequency and the $\ket{1} \rightarrow \ket{2}$ transition frequency, is also shown for each qubit along the chain in Fig.~\ref{fig:s1}C. The median value is 265 MHz.

The coherence times ($T_1$, $T_2$) for each qubit in the 1D chain are plotted in Fig.~\ref{fig:s1}D and Fig.~\ref{fig:s1}E. The median values are $T_1 = 22.2$ $\mu$s, $T_2^* = 3.2$ $\mu$s and $T_2^\text{CPMG} = 18.2$ $\mu$s. The Pauli error rates for single-qubit gates (i.e. $\pi / 2$ and $\pi$ rotations around the X or Y axis) and CZ gates are obtained from simultaneous operation of all qubits and plotted in Fig.~\ref{fig:s1}F. Here the median error rate is 0.0011 for the SQ gates and 0.0061 for the CZ gates. The conditional phase $\phi$ is characterized for each CZ gate using the technique of Floquet calibration \cite{Neill_Nature_2021, DTC_Nature_2022} and shown in Fig.~\ref{fig:s1}G. Here we find a median value of $3.142$ rad and a root-mean-square (RMS) deviation of 0.033 rad from the target value of $\pi$. Lastly, we note that Floquet calibration is also used to characterize the residual iSWAP angles of the CZ gates which are found to be very small on this quantum processor, having a median value of just $\theta = 0.003$ rad \cite{DTC_Nature_2022}.

\subsection{Gate calibration}

\begin{figure}[t!]
  \centering
  \includegraphics[width=0.5\columnwidth]{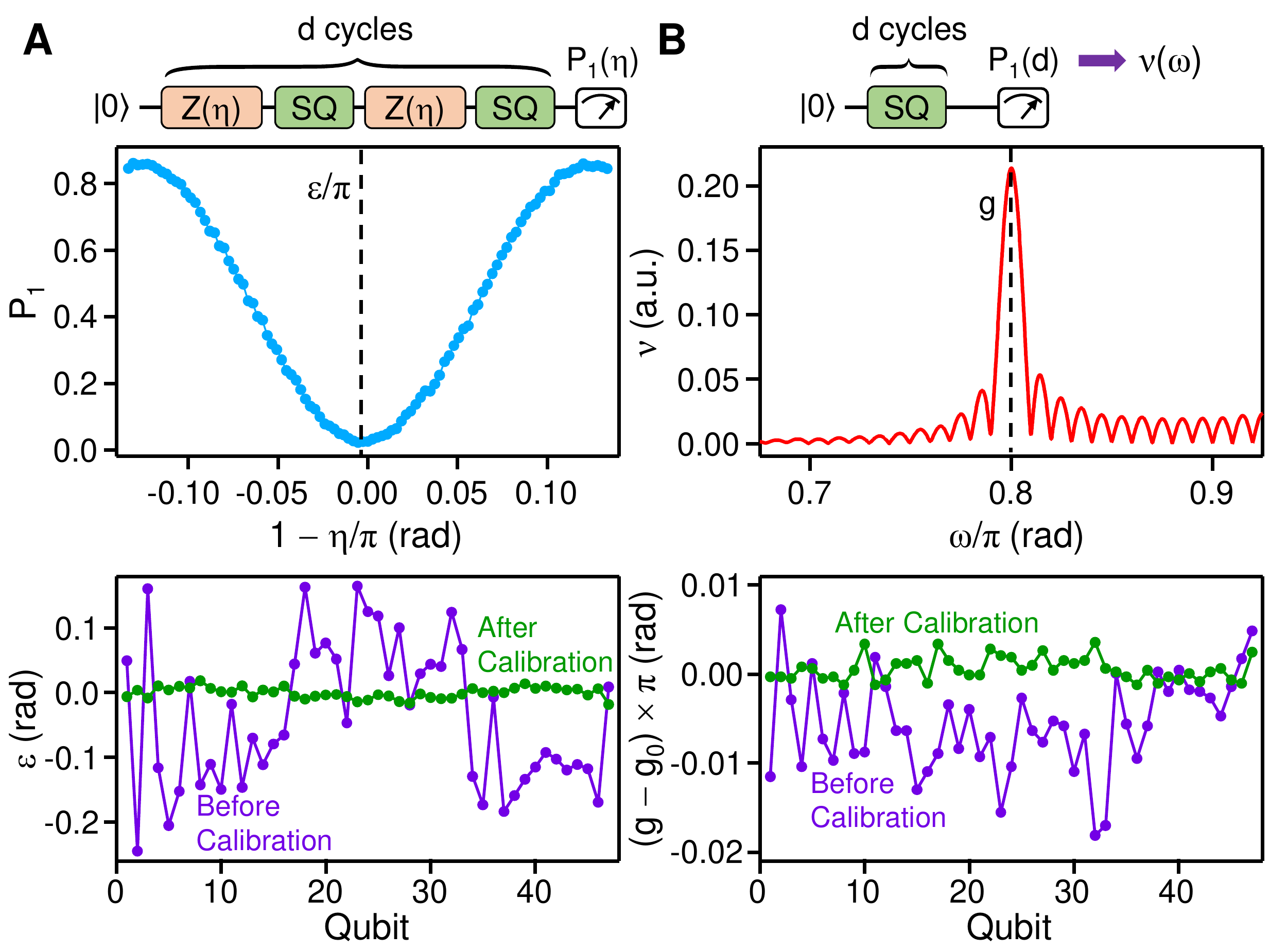} 
 \caption{Floquet calibration of single-qubit gates. (A) Top panel shows the calibration circuit used to measure the phase $\epsilon$ of a single-qubit gate, and example data set with $g = 0.8$ and averaged over $d = 12, 13, 14$. Bottom panel shows the values of $\epsilon$ for SQ gates applied to all qubits, before and after the Floquet calibration procedure. $g$ is fixed at 0.8 for all qubits. (B) Top panel shows the calibration circuit for $g$ with an example data set. Bottom panel shows the values of $g$ for SQ gates applied to all qubits, before and after the Floquet calibration procedure.}
 \label{fig:s2}
\end{figure}

For the two-qubit gates (i.e. CZ) used in this work, we employ Floquet calibration to measure and rectify coherent errors. Details on the pulse sequences used in the calibration are described in our previous publications, e.g. Section II of the Supplementary Information for Ref.~\cite{DTC_Nature_2022}. In this section, we discuss additional calibration details concerning single-qubit gates used in the experiment. Reducing coherent errors in single-qubit gates is important for resolving the energy spectra in Fig.~2 of the main text, particularly for the region where $g$ $\sim$ $0.5$.

Single-qubit gates in our experiment are realized using microwave-driven transition between $\ket{0}$ and $\ket{1}$ states of a transmon qubit. The envelope of the microwave burst is modulated by the Derivative Reduction by Adiabatic Gate (DRAG) pulse
shape to reduce leakage and phase errors \cite{Chen_PRL_2016}. The unitary for a single-gate realized using such pulse shapes may be parameterized as:
\begin{equation}
\hat{U}_\text{1q} = \begin{pmatrix}
e^{i \frac{\epsilon}{2}} \cos \left( \frac{\pi}{2} g \right) & -ie^{-i \beta} \sin \left( \frac{\pi}{2} g \right) \\
-ie^{i \beta} \sin \left( \frac{\pi}{2} g \right) & e^{-i \frac{\epsilon}{2}} \cos \left( \frac{\pi}{2} g \right)
\end{pmatrix}.
\end{equation}
$\hat{U}_\text{1q}$ has a total of three parameters, the Rabi angle $g$ and two phases $\beta$ and $\epsilon$. While $\beta$ is set by the phase of the microwave control pulse which is expected to be accurate, appreciable coherent errors in $g$ and $\epsilon$ may occur due to miscalibrated microwave power and imperfections in the DRAG pulse shape, respectively \cite{Lucero_PRA_2010, Chen_PRL_2016}.

We now describe two simple Floquet circuits that calibrate $\epsilon$ and $g$ to high precision. In the case of $\epsilon$ which has a target value of 0, we repeat an interleaved gate sequence $Z (\eta) \longrightarrow \hat{U}_\text{1q} \longrightarrow Z (\eta) \longrightarrow \hat{U}_\text{1q}$ a total of $d$ times, where $d$ is fixed (Fig.~\ref{fig:s2}A). When initialized in $\ket{0}$, the qubit returns exactly to the initial state after each Floquet cycle under the condition $1 - \eta / \pi = \epsilon / \pi$. As such, measuring the qubit excited state population $P_1$ as a function of $\eta$ and identifying its global minimum allow $\epsilon$ to be determined, as illustrated by the top panel of Fig.~\ref{fig:s2}A. Virtual $Z$ gates are then added to compensate for the non-zero value of $\epsilon$. We note that the precision of this protocol increases with the circuit depth $d$, since the width of the global minimum decreases for larger $d$. Values of $\epsilon$ across the qubit chain, obtained before and after the calibration procedure, are plotted in the bottom panel of Fig.~\ref{fig:s2}A. A significant reduction in $\epsilon$ is seen, going from a RMS value of 0.116 rad before the calibration to only 0.008 rad after the calibration.

After calibrating $\epsilon$, the Rabi angle $\theta$ is measured using another Floquet circuit shown in the top panel of Fig.~\ref{fig:s2}B. Here the single-qubit gate is repeatedly applied a total of $d$ times, with $P_1$ of the qubit measured after every gate application. The Fourier spectrum $\nu (\omega)$ of $P_1 (d)$ is then computed, where a sharp peak is seen. The peak location indicates the value of $g$, as illustrated by the example data in Fig.~\ref{fig:s2}B. Here the calibration precision also increases with $d$ since larger number of cycles lead to a narrower width of the spectral peak. Once $g$ is determined, the power of the microwave pulse is adjusted to reduce the deviation of $g$ from its target value $g_0$, $|g - g_0| \times \pi$. The calibration results across the qubit chain are plotted in the bottom panel of Fig.~\ref{fig:s2}B, where the RMS error is seen to be reduced from 0.0078 rad to 0.0014 rad through the calibration procedure.

\section{Extended data}

\subsection{Low-frequency noise response in the delocalized regimes}

\begin{figure}[t!]
  \centering
  \includegraphics[width=0.5\columnwidth]{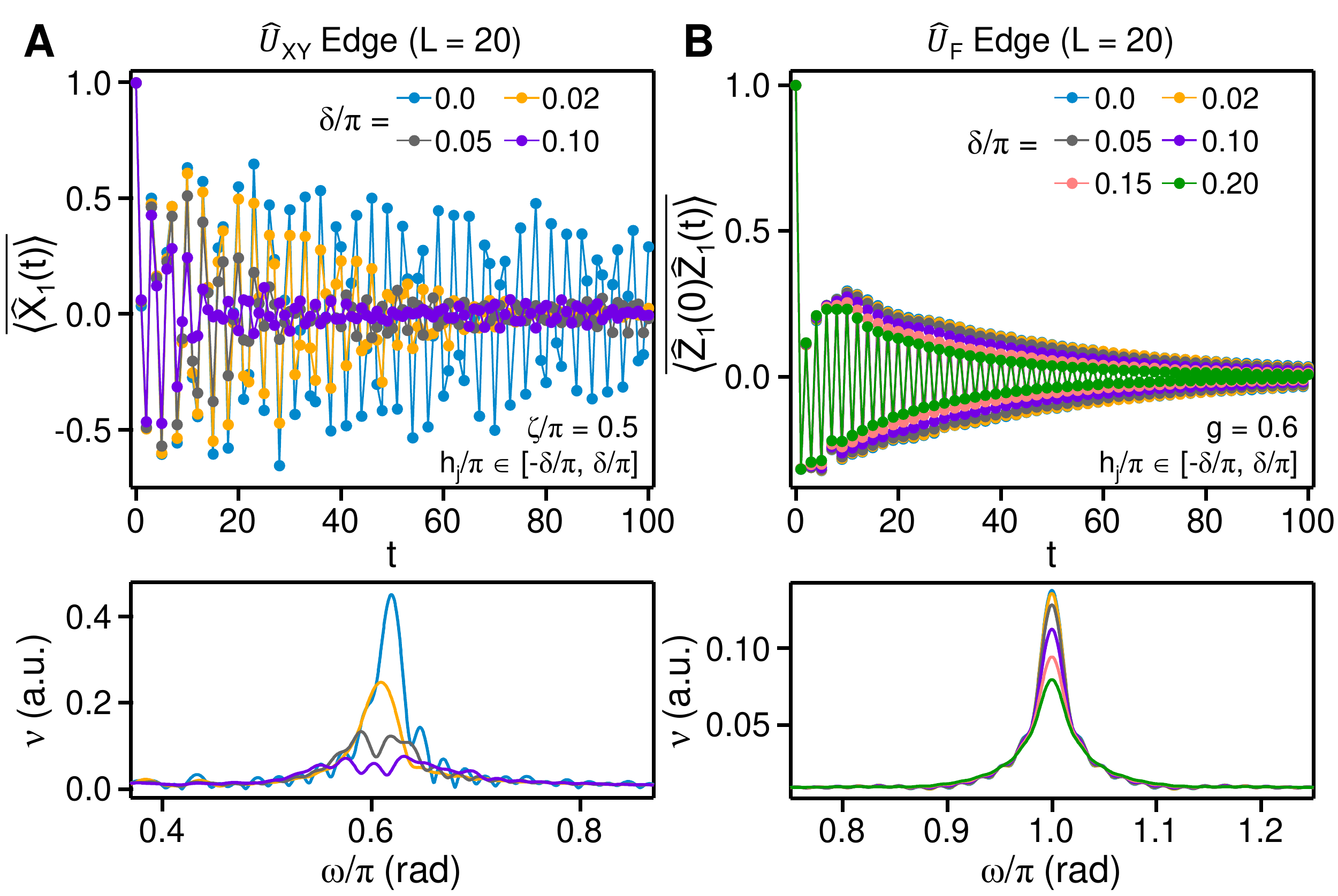} 
 \caption{Low-frequency noise response in the delocalized regimes. (A) Top panel: Disorder-averaged edge observable $\overline{\braket{\hat{X}_1 (t)}}$ for different values of disorder strength $\delta$. Here the cycle unitary is $\hat{U}_\text{XY}$ and $L = 20$. 80 disorder instances are used for averaging in each case. Bottom panel: Fourier amplitude $\nu$ as a function of frequency $\omega$, obtained from the time-domain data in the top panel. (B) Top panel: Disorder-averaged edge autocorrelator $\overline{\braket{\hat{Z}_1 (0) \hat{Z}_1 (t)}}$ for different values of disorder strength $\delta$. Here the cycle unitary is $\hat{U}_\text{F}$ and $L = 20$. 80 disorder instances are used for averaging in each case. Bottom panel: Fourier amplitude $\nu$ as a function of frequency $\omega$, obtained from the time-domain data in the top panel.}
 \label{fig:s3}
\end{figure}

In Fig.~3 of the main text, we used disorder averaging to probe the resilience to low-frequency noise and compared the behavior of the XY model ($\hat{U}_\text{XY}$) edge mode and the Majorana ($\hat{U}_\text{F}$) edge mode. The data presented there are focused on regimes where the edge modes are more spatially localized, i.e. $\zeta / \pi = 1.0$ and $g = 0.8$. Figure~\ref{fig:s3} shows the time- and frequency-domain data for the two models in their respective, more delocalized regimes ($\zeta / \pi = 0.5$ and $g = 0.6$). We observe that even in this more delocalized regime, the two models exhibit a qualitatively different behavior. The Fourier peak height $\nu_\text{max}$ decreases rapidly for the $\hat{U}_\text{XY}$ edge mode as the disorder strength $\delta$ increases. For the $\hat{U}_\text{F}$ edge mode, $\nu_\text{max}$ is much less sensitive to $\delta$ and shows little degradation for $\delta / \pi \leq 0.05$.

\subsection{Edge mode sensitivity to native device fluctuation}

\begin{figure}[t!]
  \centering
  \includegraphics[width=0.5\columnwidth]{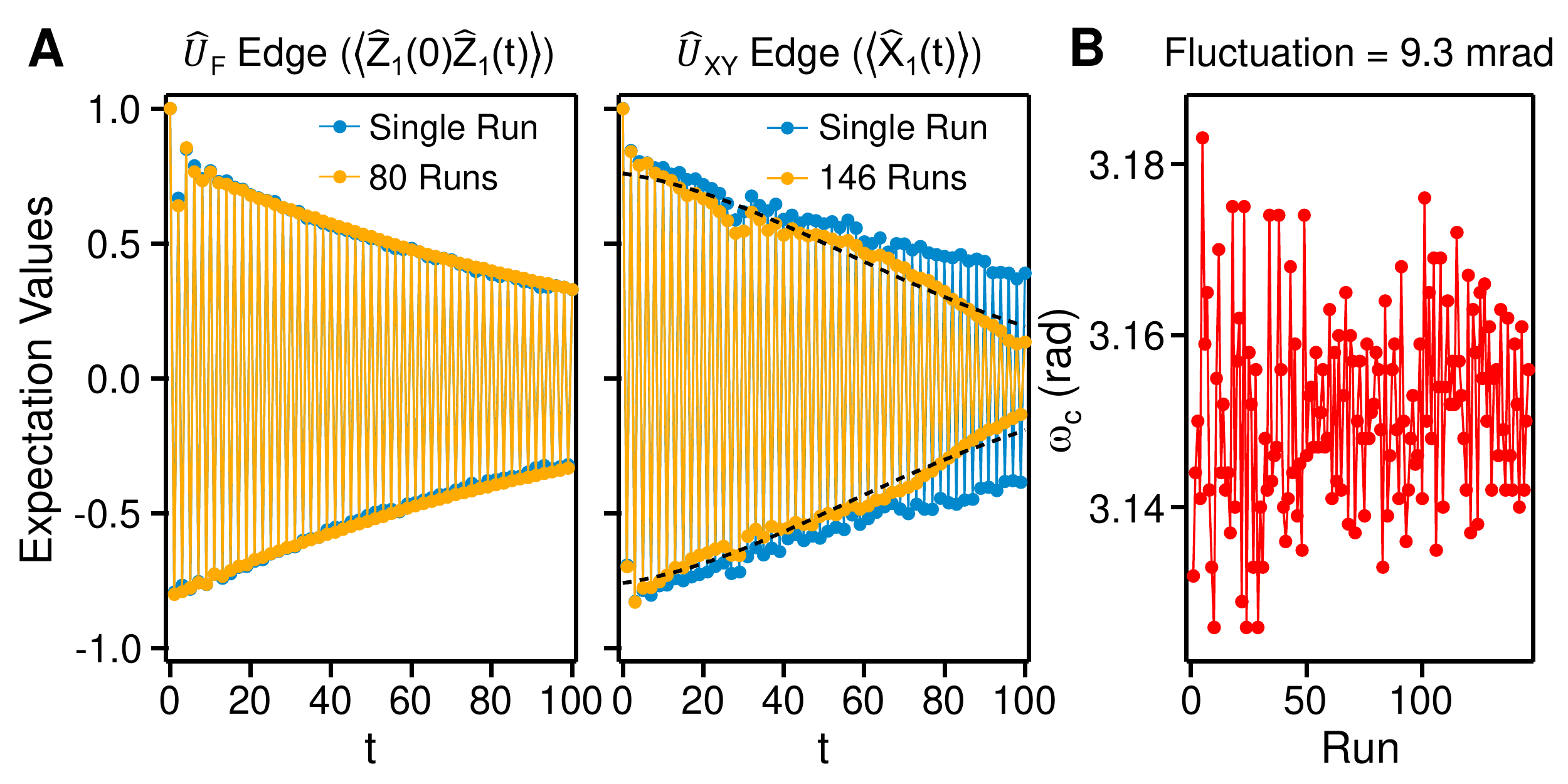} 
 \caption{Edge mode sensitivity to native device fluctuation. (A) Comparison of edge observables obtained from a single run and from the average of many nominally identical runs. Here, $g = 0.8$ for $\hat{U}_\text{F}$, $\zeta / \pi = 0.5$ for $\hat{U}_\text{XY}$ and $\delta$ is set to 0 for all cases.  Dashed lines shows the result of fitting the envelope to a functional form $\pm A_0 \exp\left( {-\frac{t}{T_\text{a}} - \frac{t^2}{T_\text{b}^2}} \right)$. (B) Quasi-energy $\omega_\text{c}$ of the $\hat{U}_\text{XY}$ edge mode, extracted from the Fourier spectrum of time-domain data in each run. An RMS fluctuation of 0.0093 rad is observed.}
 \label{fig:s4}
\end{figure}

In this section, we show experimental data that compare the impact of low-frequency noise native to the quantum device on the $\hat{U}_\text{F}$ and $\hat{U}_\text{XY}$ edge modes. As demonstrated in previous works \cite{zhang_floquet_2020}, drifts in flux bias voltages can lead to slow changes in the qubit frequencies as well as inter-qubit couplings which, in turn, lead to parameter changes in the gate unitaries.

Figure~\ref{fig:s4}A shows the time-dependent edge observables obtained from a single run of the experiment and from averaging many repetitions of the same experiment. We observe that the edge observable $\braket{\hat{Z}_1 (0) \hat{Z}_1 (t)}$ for $\hat{U}_\text{F}$ is insensitive to native device fluctuations, showing virtually identical decay rates between the two different data sets. In contrast, the edge observable $\braket{\hat{X}_1 (t)}$ shows an enhanced decay rate after averaging over runs, having a Gaussian profile characteristic of low-frequency noise \cite{Bylander_2011, Dial_PRL_2013}. A fit to the decay envelope using the functional form $\pm A_0 \exp{ \left( -\frac{t}{T_\text{a}} - \frac{t^2}{T_\text{b}^2} \right)}$, where $A_0$, $T_\text{a}$ and $T_\text{b}$ are free parameters, shows reasonable agreement with experimental data and yields a sizably shorter dephasing time $T_\text{b} = 96$ cycles (6.1 $\mu$s) compared to the relaxation time $T_\text{a} = 346$ cycles (22.1 $\mu$s). In Fig.~\ref{fig:s4}B, the quasienergy $\omega_\text{c}$ extracted from the Fourier peak location is plotted for each run of the XY model, where we observe a fluctuation of $0.0093$ rad.

\subsection{Low-frequency noise induced dephasing of MEMs for short qubit chains}

\begin{figure}[t!]
  \centering
  \includegraphics[width=1\columnwidth]{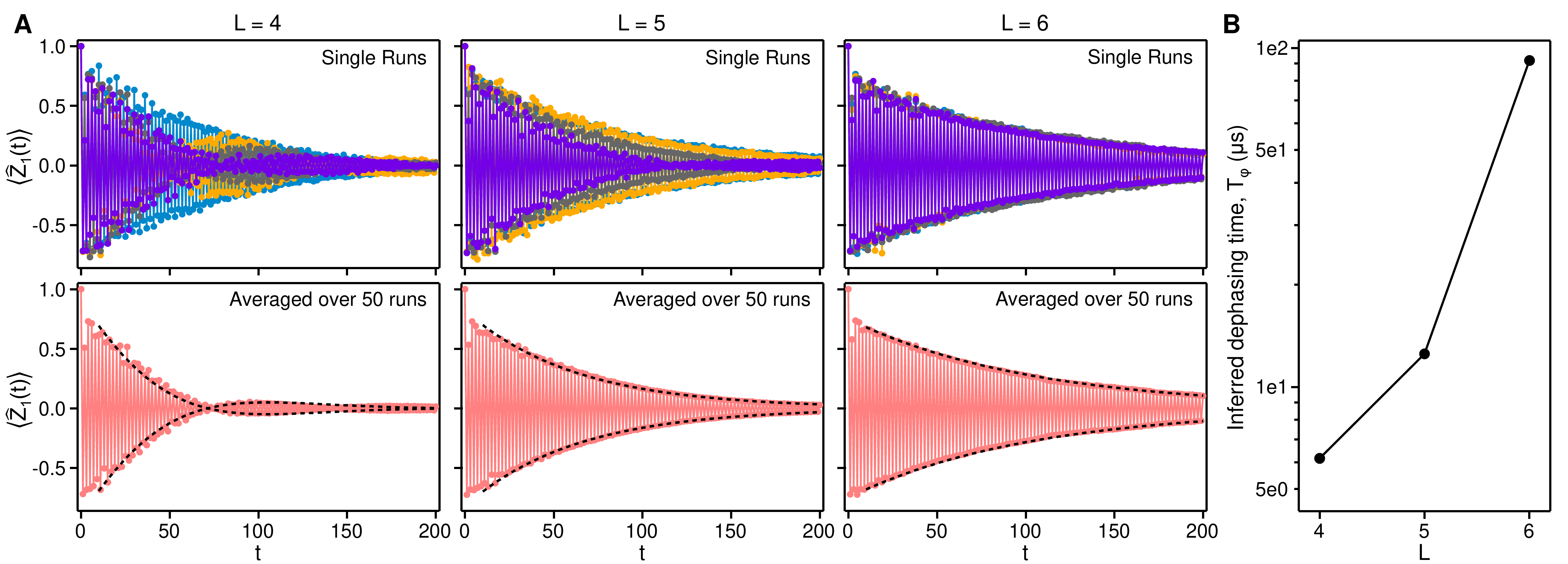} 
 \caption{Low-frequency noise induced dephasing of MEMs for short qubit chains. (A) Top panels show an edge observable $\braket{\hat{Z}_1 (t)}$ measured over 4 nominally identical experiments with $g = 0.75$ and $h_j = 0$. Data are taken for qubit chain lengths of $L = 4$, 5 and 6. Bottom panels show the same data but averaged over 50 runs. Dashed lines show fits for extracting the lifetimes (see texts for details). (B) Inferred low-frequency noise induced dephasing time, $T_\varphi$, as a function of qubit chain length $L$ (see text for analysis).}
 \label{fig:s5}
\end{figure}

As demonstrated in the main text, the MEMs are robust against low-frequency noise for sufficiently long chain such as the case of $L = 20$ studied in Fig.~3. Consequently, the low-frequency noise induced dephasing times in these long chains, $T_\varphi$, are difficult to measure since $T_\varphi$ is much larger than single-qubit $T_1$. For short qubit chains, however, the MEMs hybridize and have a quasienergy that fluctuates as a result of low-frequency noise in the system. $T_\varphi$ may be experimentally estimated in these smaller system sizes.

The top panels of Fig.~\ref{fig:s5}A show an edge observable $\braket{\hat{Z}_1 (t)}$ measured over 4 repetitions of nominally identical quantum circuits ($g = 0.75, h_j = 0$). It is seen that for the short chain $L = 4$, the subharmonic response has beat nodes which is due to the quasienergies associated with the hybridized MEMs deviating from $\pi$. The data from each repetition is also seen to be very different, which arises from the fluctuation of the MEM quasienergies due to low-frequency noise. As $L$ increases, this fluctuation is reduced as the MEMs become less hybridized and their quasienergies are more rigidly locked to $\pi$.

To estimate the lifetimes $T_\text{M}$ of the MEMs for different chain lengths, we average $\braket{\hat{Z}_1 (t)}$ over 50 repetitions and plot the results in the lower panels of Fig.~\ref{fig:s5}A. Here we see clearly slower decay as $L$ increases. To estimate $T_\text{M}$, we then fit the decay envelope of $\braket{\hat{Z}_1 (t)}$ associated with each $L$ to a functional form $\pm \text{Re} \left[ A_0 \exp{ \left( -\frac{t}{T_\text{M}} - i\omega_\text{d} t \right)} \right]$ where $A_0$ and $T_\text{M}$ are free parameters. $\omega_\text{d}$ is an additional fitting parameter for $L = 4$ to account for the deviation of the MEM quasienergies from $\pi$ and set to 0 for $L = 5$ and $L = 6$. 

Lastly, to estimate $T_\varphi$ at these chain lengths, we assume that $T_\phi (L = 47) \gg T_\phi (L = 4, 5 \text{ or } 6)$ and that the MEMs have comparable lifetimes in the absence of dephasing for $L = 4, 5, 6 \text{ or } 47$ (a reasonable assumption given that the quasienergy gap between the MEMs and the bulk modes is comparable between the four cases). Under these assumptions, $T_\phi$ for small $L$ is then directly related to the decrease in $T_\text{M}$ and given by the relation $\frac{1}{T_\phi} = \frac{1}{T_\text{M}} - \frac{1}{T_\text{M} (L = 47)}$, where $T_\text{M} (L = 47)$ is the MEM lifetime measured at $g = 0.75, L = 47$ (see Fig.~\ref{fig:s8}B below). The inferred value of $T_\phi$ is shown as a function of $L$ in Fig.~\ref{fig:s5}B, where a sharp increase in $T_\phi$ is seen as $L$ increases. These results demonstrate that the low-frequency noise sensitivity of MEMs is only suppressed in qubit chains of sufficient lengths.

\subsection{Multi-qubit correlators for the right edge mode}

\begin{figure}[t!]
  \centering
  \includegraphics[width=0.5\columnwidth]{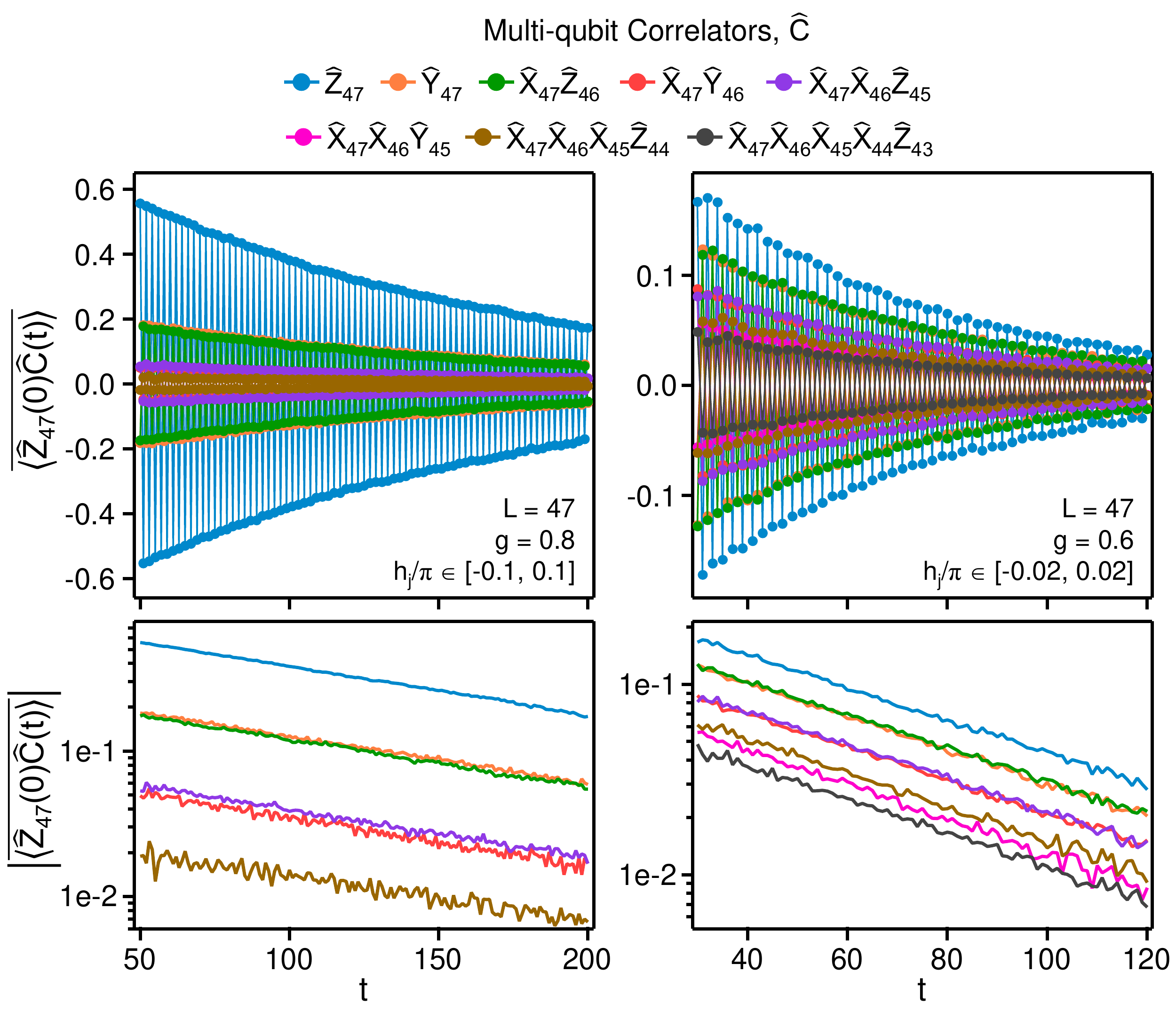} 
 \caption{Multi-qubit correlators for the right edge mode. Upper panels: Time-dependent data for 8 different multi-qubit correlators obtained on the right end of the 47-qubit chain. Similar to the left edge mode shown in the main text, a small local field disorder is added in each case. The plotted data correspond to averaging over 10 (12) instances of disorder realizations as well as random initial states, for $g = 0.8$ ($g = 0.6$). Lower panels: Absolute values of the correlators as a function of time, shown with a log vertical scale.}
 \label{fig:s6}
\end{figure}

The multi-qubit correlators taken at the right end of the 47-qubit chain are shown in Fig.~\ref{fig:s6}. Their behavior is very similar to the data on the left end (shown in Fig.~4A of the main text), namely an identical decay for all correlators at a fixed value of $g$ is observed.

\subsection{Multi-qubit correlators in the presence of constant $z$-field}

\begin{figure}[t!]
  \centering
  \includegraphics[width=0.5\columnwidth]{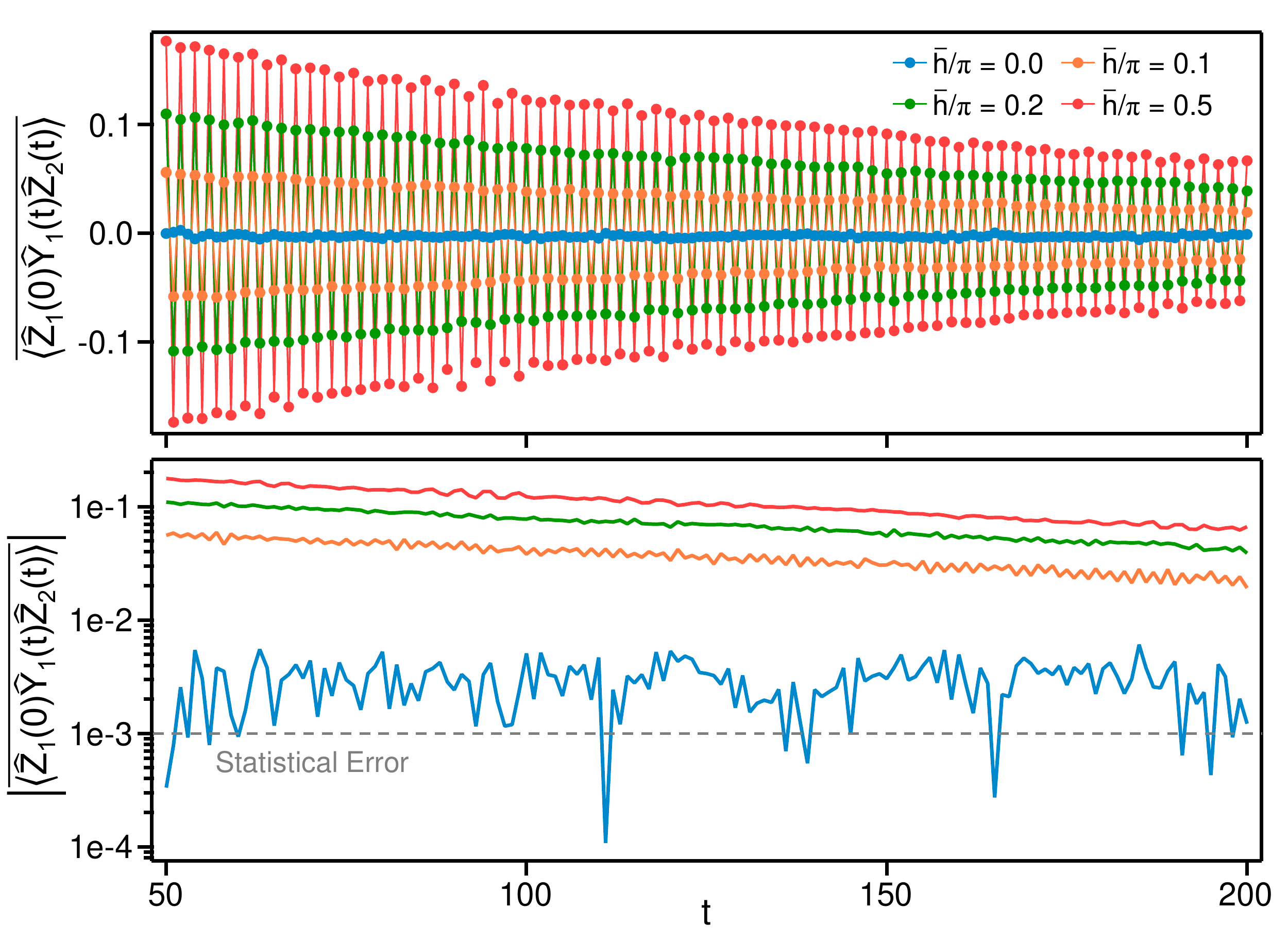} 
 \caption{Multi-qubit correlators in the presence of constant $Z$ field. Top panel: Experimentally measured $\overline{\braket{\hat{Z}_1 (0) \hat{Y}_1 (t) \hat{Z}_2 (t)}}$ for different average local fields $\overline{h}$, such that $h_j \in [\overline{h} - 0.1, \overline{h} + 0.1]$. Bottom panel: Absolute values $\left| \overline{\braket{\hat{Z}_1 (0) \hat{Y}_1 (t) \hat{Z}_2 (t)}} \right|$ as a function of $t$. Dashed line denotes statistical errors estimated from the number of single-shot measurements.}
 \label{fig:s7}
\end{figure}

The observation that any multi-qubit operator overlapping with conserved quantities such as the MEMs acquires a slow decay at late times (with an amplitude proportional to the overlap) allowed us to reconstruct the expansion of the MEMs in the Pauli basis in Fig.~4B of the main text. Interestingly, this observation is not restricted to integrable dynamics (i.e. $h_j \approx 0$ in our model) and applies to non-integrable dynamics as well. As a demonstration, the upper panel of Fig.~\ref{fig:s7} shows the time-dependent expectation values of a particular operator $\hat{Y}_1 \hat{Z}_2$ in the presence of a constant, non-zero integrability-breaking field $\overline{h}$. 

At $\overline{h} = 0$, the operator has no overlap with the MEMs and nearly zero expectation values that are comparable to the statistical uncertainties of the measurements (see lower panel of Fig.~\ref{fig:s7}). As $\overline{h}$ increases, $\hat{Y}_1 \hat{Z}_2$ develops non-zero overlap with the MEMs and shows slowly-decaying subharmonic oscillations with an envelope that changes with $\overline{h}$. Notably, the decay rate does not depend on the overall value of $\overline{h}$. When combined with other multi-qubit operator measurements (which we leave to future works), the Pauli expansion of MEMs in the presence of constant $Z$ fields should in principle be possible to construct experimentally using a similar strategy as Fig.~4B of the main text. 

While the response here can still be theoretically approximated (Section \ref{sec:perturb} below), these data nevertheless indicate the feasibility of applying our experimental protocol to other non-integrable dynamics that are more difficult to simulate classically.

\subsection{Edge mode lifetime as a function of $g$}

\begin{figure}[t!]
  \centering
  \includegraphics[width=0.5\columnwidth]{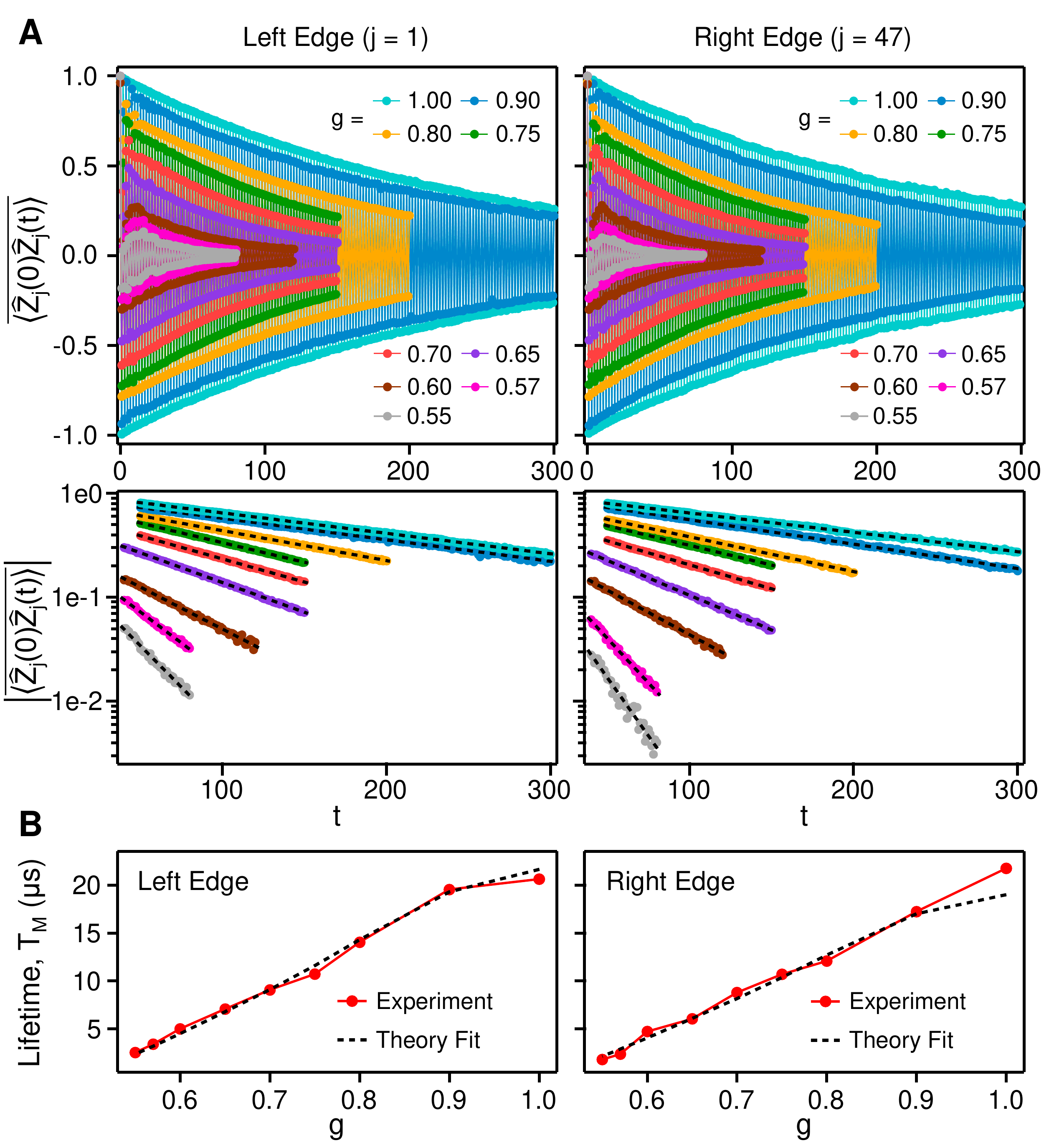} 
 \caption{Edge mode lifetime as a function of $g$. (A) Upper panels: Auto-correlators of the edge qubits, $\overline{\braket{\hat{Z}_1 (0) \hat{Z}_1 (t)}}$ and $\overline{\braket{\hat{Z}_{47} (0) \hat{Z}_{47} (t)}}$, obtained for different values of $g$. Lower panels: Absolute values of the auto-correlators shown for late times and plotted on a log vertical scale. Each dashed line shows fit to an exponential decay, which allows the lifetime $T_\text{M}$ to be extracted for each value of $g$. (C) The extracted $T_\text{M}$ as a function of $g$ for both the left and the right edge modes. Dashed lines show fits to theory.}
 \label{fig:s8}
\end{figure}

Figure~\ref{fig:s8}A shows the leading order terms in the $\hat{U}_\text{F}$ edge operators, $\overline{\braket{\hat{Z}_1 (0) \hat{Z}_1 (t)}}$ and $\overline{\braket{\hat{Z}_{47} (0) \hat{Z}_{47} (t)}}$, for different values of $g$. In each case, the local field disorder is kept at a sufficiently low level such that the observed decay is dominated by external decoherence effects. More specifically, we have chosen the disorder strength $\delta / \pi$ to be 0.10 for $g \geq 0.75$, 0.02 for  $g \geq 0.60$ and 0 for  $g < 0.60$. The late-time decay for each dataset is seen to be exponential and fitted to a functional form $A_0 \exp{\left(-\frac{t}{T_\text{M}} \right)}$, where $T_\text{M}$ is the lifetime of the edge mode.

The dependence of $T_\text{M}$ on $g$ for each edge mode is shown in Fig.~\ref{fig:s8}B. The experimental data are then fitted against the theoretical results described in Section~\ref{app:dissipation}, where two free parameters $\gamma_\text{d}$ (single-qubit relaxation rate) and $\gamma_\phi$ (single-qubit dephasing rate) are used in the fitting. The best-fit results are $1/\gamma_\text{d} = 21.6$ $\mu$s (19.0 $\mu$s) and $1/\gamma_\phi = 9.3$ $\mu$s (8.5 $\mu$s) for the left (right) edge mode. The values of $1/\gamma_\text{d}$ obtained here are in good agreement with typical single-qubit $T_1$ separately characterized in Fig.~\ref{fig:s1}A. 

While the phenomenological model considered here illustrates a good agreement with data, a physics-based model with more realistic assumptions that takes effects such as spatial non-uniformity of decoherence and decays into account as well residual ``many-body'' couplings between physical qubits \cite{Berke2022} can provide deeper insight into the unresolved discrepancies between data and experiment.

\subsection{Comparison between edge mode and single-qubit Rabi experiments}

\begin{figure}[t!]
  \centering
  \includegraphics[width=1\columnwidth]{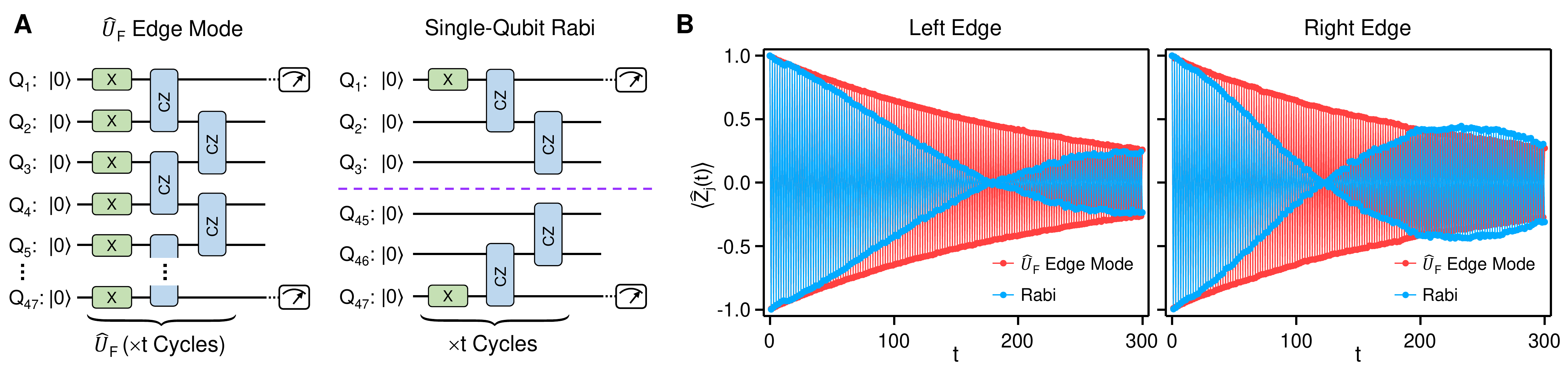} 
 \caption{Comparison between edge mode and single-qubit Rabi experiments. (A) Circuits for $\hat{U}_\text{F}$ edge mode (left panel) and a single-qubit Rabi experiment (right panel). For the $\hat{U}_\text{F}$ edge mode experiment, $g$ is set to 1 and CZ gates are used in place of $\sqrt{\text{ZZ}}$ gates. For the single-qubit Rabi experiment, an $X$ gate is applied to only the leftmost ($Q_1$) and rightmost ($Q_{47}$) qubits within each cycle, and CZ gates are only applied between three qubit on the very left ($Q_1$ through $Q_3$) and three qubits on the very right ($Q_{45}$ through $Q_{47}$). (B) Comparison of $\braket{\hat{Z}_j (t)}$ between the two experiments, shown both for the left edge ($j = 1$) and right edge ($j = 47$).}
 \label{fig:s9}
\end{figure}

The stability of the Majorana edge modes against various perturbations demonstrated in our work raises the tantalizing prospect of their future applications to engineering noise-resilient qubits. Although a detailed discussion on the possible schemes of encoding qubits into such edge modes is outside the scope of this work, we demonstrate the potential benefits of Majorana edge modes by comparing them with a simple Rabi oscillation experiment. The circuits used for this comparison are shown in Fig.~\ref{fig:s9}A. Here the Majorana edge modes are measured with $g = 1$ and the single-qubit Rabi experiment is done by applying $X$ gates only to the leftmost and rightmost qubits. To ensure that the execution time per cycle is equal in both cases, we have also added ``padding'' in the form of four additional CZ gates for the Rabi experiment (which effectively act as identity gates since there is at most one excitation within the system).

The observables $\braket{\hat{Z}_j (t)}$ for the two edge qubits are shown in Fig.~\ref{fig:s9}B. Here we see a clear difference between the single-qubit Rabi experiment and behavior of the Majorana edge modes: For the Rabi experiment, coherent errors in the $X$ gate (which can arise from, e.g. miscalibrated Rabi power or imperfect DRAG coefficients for the microwave pulses \cite{Chen_PRL_2016}) accumulate over cycles and cause beating in $\braket{\hat{Z}_j (t)}$ at long times. On the other hand, the stable quasienergy ($= \pi$) of the Majorana edge modes allows steady oscillation between $\ket{0}$ and $\ket{1}$ states of the edge qubits, despite the imperfection of the $X$ gates. These results are a preliminary indication that the phase structure of the Majorana edge modes could be harnessed to improve quantum information processing of superconducting qubits.

\subsection{MEMs in the $g < 0.5$ regime}

\begin{figure}[t!]
  \centering
  \includegraphics[width=1\columnwidth]{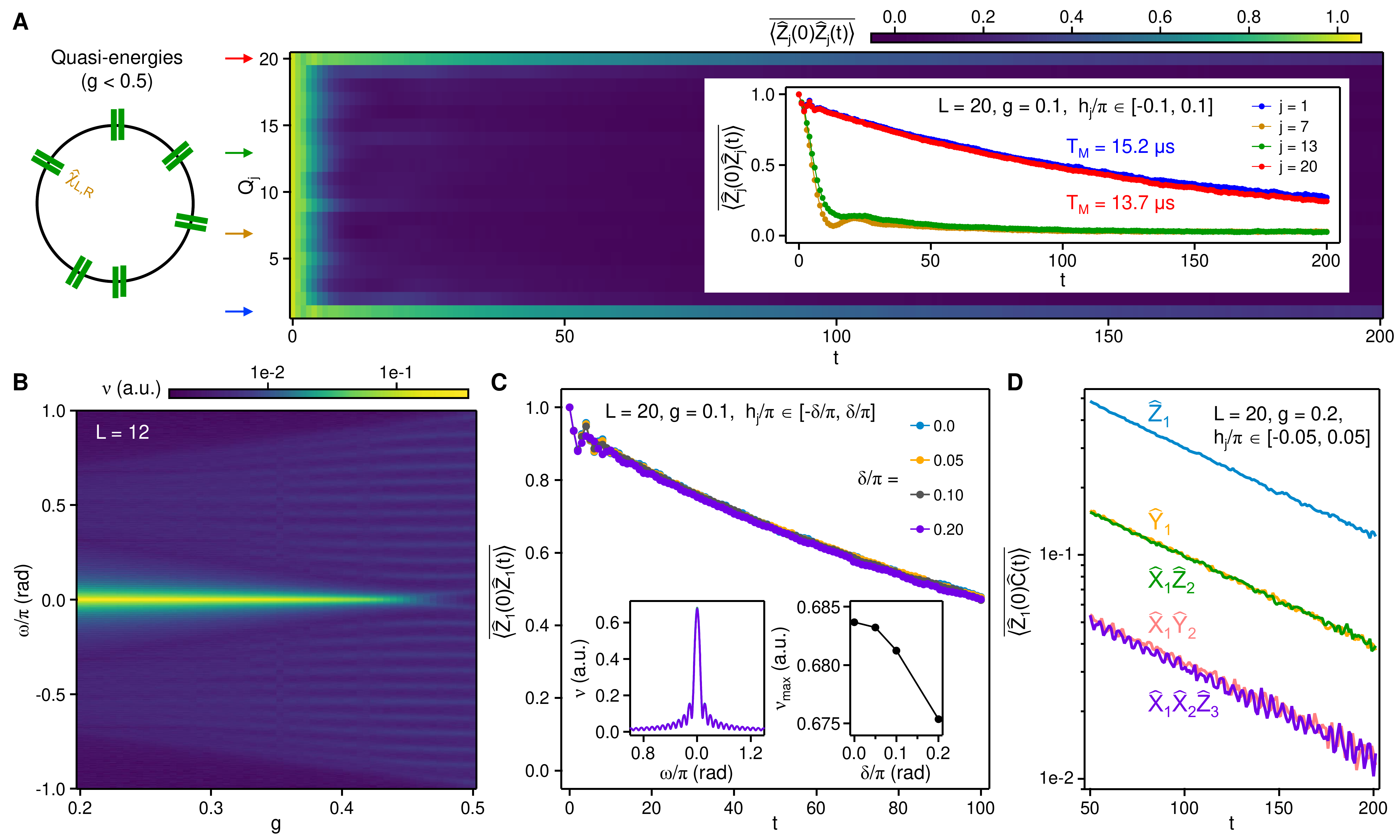} 
 \caption{MEMs in the $g < 0.5$ regime. (A) Left panel: Schematic illustration of the quasienergy spectrum for the $g < J = 0.5$ regime of the kicked Ising model. Eigenstates of $\hat{U}_\text{F}$ are shown on a unit circle according to their quasienergies. Right panel: $\overline{\braket{\hat{Z}_j (0) \hat{Z}_j (t)}}$ as a function of $t$ and qubit location $Q_j$, averaged over 20 instances of initial random product states and local field disorder $h_j / \pi \in [-0.1, 0.1]$. The system size is $L = 20$ with $g = 0.1$. Inset shows $\overline{\braket{\hat{Z}_j (0) \hat{Z}_j (t)}}$ as a function of $t$ for the two edge qubits ($j = 1$ and $j = 20$) and two qubits in the bulk ($j = 7$ and $j = 13$). (B) Experimentally measured quasienergy spectrum as a function of $g$. Each vertical column $\nu (\omega)$ is obtained by measuring $\braket{\hat{Z}_1 (t)}$ up to $t = 200$ and then Fourier transforming the time-domain signal, same as Fig.~2 of the main text. Here $L = 12$ and $h_j = 0$. (C) $\overline{\braket{\hat{Z}_1 (0) \hat{Z}_1 (t)}}$ as a function of $t$, averaged over 30 instances of random initial product states and local field disorder $h_j / \pi \in [-\delta / \pi, \delta / \pi]$. Data are shown for four values of the disorder strength $\delta$. Left inset shows the Fourier spectra $\nu (\omega)$ of the four time-domain signals. Right inset shows the maximum Fourier amplitude $\nu_\text{max} = \text{Max}[\nu (\omega)]$ as a function of $\delta$. (D) Correlators $\overline{\braket{\hat{Z}_1 (0) \hat{C}(t)}}$ for $g = 0.2$ and $h_j / \pi \in [-0.05, 0.05]$, where a total of 5 different operators $\hat{C}$ are measured. Here the data are averaged over 18 disorder realizations and initial random product states.}
 \label{fig:s10}
\end{figure}

In the main text, we have primarily focused on the $g \geq 0.5$ regimes of the Floquet system, with $J$ fixed to a value of 0.5. In this section, we first show experimental data in the $g < 0.5$ regime and demonstrate that the noise resilience observed for the MEMs applies to this regime as well. The left panel of Fig.~\ref{fig:s10}A shows the many-body spectrum of the system in the integrable limit (i.e. $h_j = 0$) with $g < J = 0.5$. Here the eigenspectrum of $\hat{U}_\text{F}$ is doubly degenerate in the long chain limit ($L = \infty$) and each eigenstate has a partner state of equal quasienergy. Similar to the MEMs in the $g >0.5$ regime, The MEMs $\hat{\chi}_{\text L, R}$ in this regime also induce transition between the paired eigenstates in the spectrum. 

The right panel of Fig.~\ref{fig:s10}A shows the disorder and initial state averaged auto-correlators $\overline{\braket{\hat{Z}_j (0) \hat{Z}_j (t)}}$ as a function of qubit location $j$ and the number of Floquet cycles $t$ in a $L = 20$ chain. Similar to the $g > 0.5$ regime measured in Fig.~1 of the main text, we observe dramatically slower decay rates for the edge qubits compared to qubits in the bulk. The lifetimes of the MEMs in this regime, $T_\text{M}$, are also shown in Fig.~\ref{fig:s10}A and found to be comparable to those observed in the $g > 0.5$ regime. We make two additional remarks: (1) The subharmonic oscillation observed for $g > 0.5$ is no longer present in this part of the phase diagram. This is because $\hat{\chi}_{\text L, R}$ commutes, instead of anticommutes, with $\hat{U}_\text{F}$ and therefore is conserved after every application of $\hat{U}_\text{F}$ instead of acquiring a negative sign. (2) We have chosen a smaller disorder $h_j$ compared to Fig.~1 of the main text since at large disorder, many-body localization (MBL) may occur at this part of the phase diagram and slow down the decay of the bulk qubits as well \cite{ising_mbl_2022}. We leave detailed studies of MBL and its interplay with the MEMs as subjects of future work.

The single-particle quasienergy spectrum of $\hat{U}_\text{F}$ in the integrable limit ($h_j = 0$) is obtained using the same technique as Fig.~2 of the main text and shown in Fig.~\ref{fig:s10}B for a chain length of $L = 12$. At $g < 0.45$, we observe a dominant peak at zero frequency $\omega = 0$, corresponding to the MEMs in this regime which have a quasienergy close to 0 and are referred to as 0-MEMs. At $g$ close to 0.5, the 0-MEMs hybridize and split in quasienergy, similar to the behavior of the $\pi$-MEMs in the $g > 0.5$ regime.

The resilience of the 0-MEMs against low-frequency noise is tested using the same technique as Fig.~3 of the main text. In Fig.~\ref{fig:s10}C, we show the initial-state and disorder averaged $\overline{\braket{\hat{Z}_j (0) \hat{Z}_j (t)}}$ for four values of the disorder strength $\delta$ over which $h_j$ is drawn, $h_j / \pi \in [-\delta / \pi, \delta / \pi]$. We observe that the decay rate of the $\overline{\braket{\hat{Z}_j (0) \hat{Z}_j (t)}}$ is nearly unchanged by $\delta$. This is further elucidated by Fourier-transforming $\overline{\braket{\hat{Z}_j (0) \hat{Z}_j (t)}}$ (shown in the left inset of Fig.~\ref{fig:s10}C) and plotting the maximum Fourier peak height $\nu_\text{max}$ as a function of $\delta$ (shown in the right inset of Fig.~\ref{fig:s10}C). A merely 1\% decay is observed for $\nu_\text{max}$ as $\delta / \pi$ is increased from 0 to 0.2, indicating a similar robustness against low-frequency noise as the $\pi$-MEMs in the $g > 0.5$ regime.

Lastly, we show that similar to the $\pi$-MEMs, multi-qubit Pauli operators overlapping with the 0-MEMs also have identical decay rates. Figure~\ref{fig:s10}D shows experimental measurements of five different operators $\overline{\braket{\hat{Z}_1 (0) \hat{C}(t)}}$ where $\hat{C}$ is a multiqubit operator including up to three qubits. The late-time ($t > 50$) values of these operators have amplitudes that depend on their respective overlap with the $\hat{\chi}_{\text L, R}$ but a uniform decay rate. Together, these results indicate that all experimental findings in the main text apply to the $g < 0.5$ regime of the phase diagram as well.

\subsection{MEMs in the Trotterized transverse Ising model}

\begin{figure}[t!]
  \centering
  \includegraphics[width=1\columnwidth]{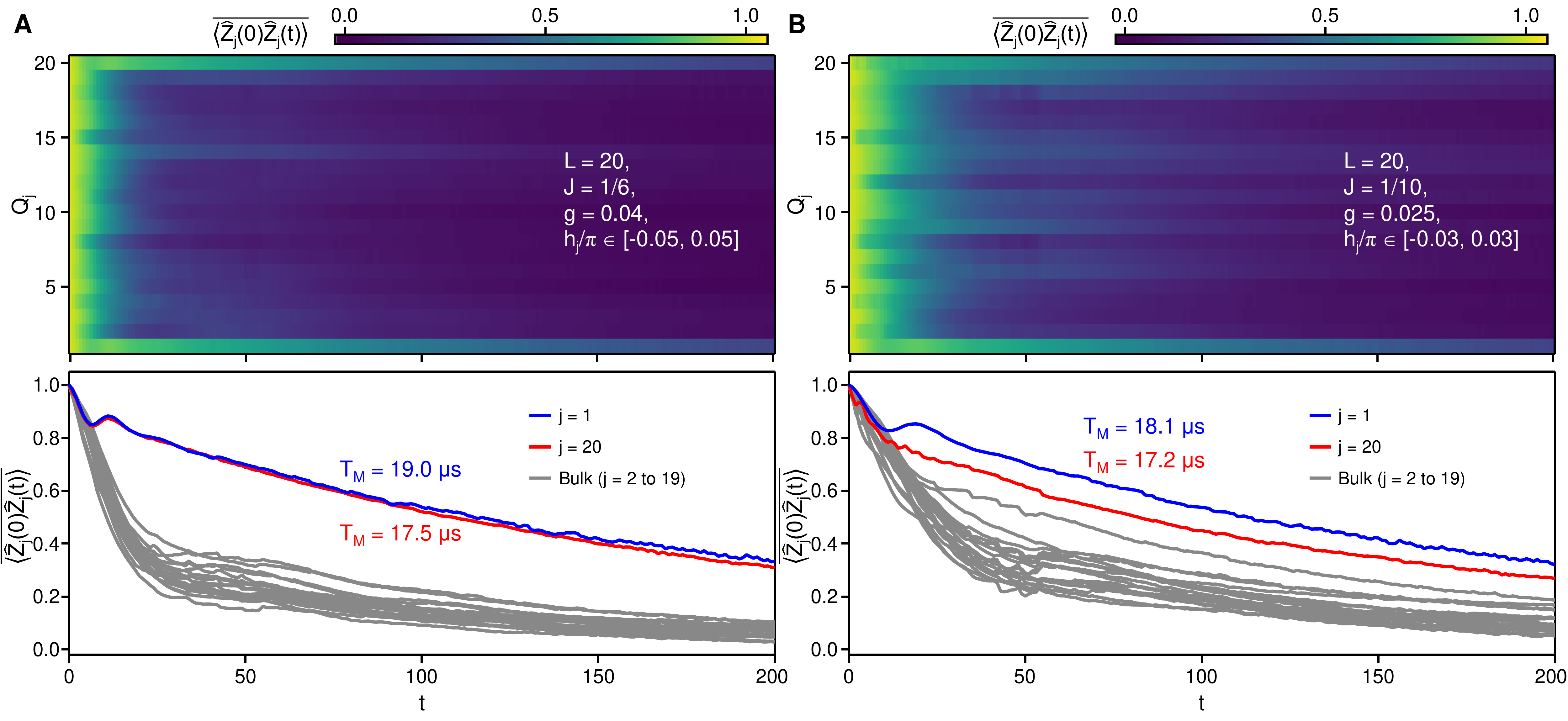} 
 \caption{MEMs in the Trotterized transverse Ising model. (A) Top panel: $\overline{\braket{\hat{Z}_j (0) \hat{Z}_j (t)}}$ as a function of $t$ and qubit location $Q_j$, averaged over 40 instances of initial random product states and local field disorder $h_j / \pi \in [-0.05, 0.05]$. The system size is $L = 20$ with $g = 0.04$ and $J = 1/6$. Bottom panel: $\overline{\braket{\hat{Z}_j (0) \hat{Z}_j (t)}}$ as a function of $t$ for the two edge qubits ($j = 1$ and $j = 20$) and all qubits in the bulk ($j = 2$ to 19). (B) $\overline{\braket{\hat{Z}_j (0) \hat{Z}_j (t)}}$ as a function of $t$ and qubit location $Q_j$, averaged over 40 instances of initial random product states and local field disorder $h_j / \pi \in [-0.03, 0.03]$. The system size is $L = 20$ with $g = 0.025$ and $J = 1/10$. Bottom panel: $\overline{\braket{\hat{Z}_j (0) \hat{Z}_j (t)}}$ as a function of $t$ for the two edge qubits ($j = 1$ and $j = 20$) and all qubits in the bulk ($j = 2$ to 19).}
 \label{fig:s11}
\end{figure}

The experiments in the main text and the majority of the SM are conducted with the strongly driven kicked Ising model. Although extending our findings to the time-independent transverse Ising model is left mainly as a subject of future study, we nevertheless present preliminary results toward this direction. Here, we have calibrated tunable CPHASE gates on a chain of $L = 20$ qubits (see Ref.~\cite{DTC_Nature_2022} for details on gate implementation) which allow the strength of the $ZZ$ interaction, $J$, to be freely adjusted. We then measure the disorder and initial state averaged autocorrelators $\overline{\braket{\hat{Z}_j (0) \hat{Z}_j (t)}}$ for $g = 0.04, J = 1/6$ (Fig.~\ref{fig:s11}A) and $g = 0.025, J = 1/10$ (Fig.~\ref{fig:s11}B), respectively. The smaller magnitudes of these parameters bring the time-evolution closer to that of a time-independent transverse Ising model, with Fig.~\ref{fig:s11}B having a smaller Trotter step than Fig.~\ref{fig:s11}A. Here the disorder strength in $h_\text{j}$ is chosen to be small enough to avoid MBL in the bulk of the qubit system but still comparable to the native noise level of typical superconducting qubit processors. Compared to qubits in the bulk, we again observe notably slower decay in the edge qubits with lifetimes $T_\text{M}$ comparable to single-qubit $T_1$. These experiments are therefore indications that the noise resilience of the edge qubits likely applies to the time-independent transverse Ising model as well.

\section{Edge modes in an integrable chain}\label{app1}

In this part of the Supplementary Material, we discuss the non-equilibrium phase diagram of the kicked transverse-field Ising model, its spectral properties, and Majorana-like edge modes~\cite{DuttaPRB13,Mitra19,LeroseInt21}. 

The Floquet operator of the kicked transverse-field Ising model is given by:
\begin{equation}
\label{eq:KIC}
\hat{U}_{\rm F,0}=e^{-\frac{i\pi J}{2}\sum_{i=1}^{L-1} \hat{Z}_i \hat{Z}_{i+1}} e^{-\frac{i\pi g}{2}\sum_{i=1}^L \hat{X}_i}.
\end{equation}
Here and below $\hat{X}_i, \hat{Y}_i, \hat{Z}_i$ denote Pauli matrices acting on qubit $i$. The model can be solved by mapping the generators of the unitary gates to bilinear forms of fermionic creation/annihilation operators $c_j^\dagger$, $c_j$, via a Jordan-Wigner transformation: 

\begin{equation}
    \hat\sigma^-_j
    = \prod_{i=1}^{j-1} e^{i\pi c^\dagger_i c_i} c^\dagger_j
    , \qquad
    \hat{X}_j
    = 1-2c^\dagger_j c_j = e^{i\pi c^\dagger_j c_j},
\end{equation}
where $\hat\sigma^{\pm}_j=\frac 1 2 (\hat{Y}_j \pm i \hat{Z}_j)$.
The operators $c_j$, $c_j^\dagger$ defined above satisfy the canonical fermionic algebra
\begin{equation}
\{ c_i, c_j \} = 0 , \qquad
\{ c_i, c^\dagger_j \} = \delta_{ij} .
\end{equation}
In the fermionic language, the model (\ref{eq:KIC}) takes the following form: 
\begin{equation}\label{eq:Ffermions}
\hat{U}_{\rm F,0}=\prod_{j=1}^{L-1}
e^{
-\frac{i\pi J}{2}
(c^\dagger_j+c_j) (c_{j+1} -
 c^\dagger_{j+1})
}
  \prod_{j=1}^L
  e^{-\frac{i \pi g}{2} (c_j c^\dagger_j-c^\dagger_jc_j)
}
\end{equation}

As a next step, we introduce Majorana operators:
 \begin{equation}
    a_{2j-1}=i(c_j^\dagger-c_j),
      \qquad
      a_{2j}=c_j+c_j^\dagger,
 \end{equation}
 with $\{a_m,a_n\}=2\delta_{mn}$. The Majorana operators in terms of spin operators read: 
\begin{equation}\label{eq:maj_spins}
a_{2j-1}= \left[\prod_{i=1}^{j-1}\hat{X}_i \right] \hat{Z}_j,\;\;   a_{2j}=\left[\prod_{i=1}^{j-1} \hat{X}_i \right] \hat{Y}_j.
\end{equation} 
In terms of the Majorana operators, the Floquet operator becomes:
\begin{equation}\label{eq:Fmajoranas}
\hat{U}_{\rm F,0}=\prod_{j=1}^{L-1}
e^{
\frac{\pi J}{2}
a_{2j} a_{2j+1}
}
\prod_{j=1}^{L}
e^{
 \frac{ \pi g}{2} a_{2j-1} a_{2j}
}\end{equation}

\subsection{Quasienergy spectrum and phase diagram}

In this Subsection, we briefly describe the non-equilibrium phase diagram of the above model~\cite{DuttaPRB13}. Even though we are mainly interested in the behavior of edge modes, to establish the phase diagram we consider periodic boundary conditions. 
Then, the quasienergy spectrum $\pm \phi_k$ as a function of the momentum $k$ is given by
\begin{equation}\label{eq:phi_k}
\cos \phi_k=\cos \pi J \cos \pi g + \sin \pi J \sin \pi g \cos k .
\end{equation}
The two quasienergy bands are generally separated by a gap, which closes at $k=0$ or $k=\pi$ when $J=g$ or $J=1-g$. The gap closing signals a phase transition between distinct topological Floquet phases, some of which feature edge modes with $\phi^e=0,\pi$ (arising for $g < J$ and $g > 1-J$, respectively~\cite{DuttaPRB13}). We note in passing that the spectrum of bulk eigenstates in a finite-size chain with open boundary conditions can be found by solving the problem of scattering of the states (\ref{eq:phi_k}) at the boundary, which yields the quantization condition. 

\subsection{Edge modes}

To analyze the existence of edge modes and their structure, we first write the eigenvalue equation for the Floquet operator. We look for the eigenmodes in the form
$$
\sum_j \psi_{2j-1} a_{2j-1}+\psi_{2j} a_{2j}. 
$$
For our purposes, it is convenient to write the eigenvalue equation in terms of a transfer matrix:
\begin{equation}
    \begin{pmatrix}
    \psi_{2j+1} \\
    \psi_{2j+2}
    \end{pmatrix}
    =
    T_\phi
    \begin{pmatrix}
    \psi_{2j-1} \\
    \psi_{2j}
    \end{pmatrix},
\end{equation}
where
\begin{equation}
    T_\phi=
    \frac{1}{\sin(\pi J)}
    \begin{pmatrix}
    \sin(\pi g)e^{-i\phi}
    &
    \cos(\pi g)e^{-i\phi}-\cos(\pi J)
    \\
    \cos(\pi g)e^{-i\phi}- \cos(\pi J)
    &
    \quad\frac{e^{i\phi}}{\sin(\pi g)} - 2 \cos(\pi J) \cot(\pi g)
    + e^{-i\phi} \cos(\pi g)\cot(\pi g)
    \end{pmatrix}.
\end{equation}

This matrix has a unit determinant, $\det T_\phi = 1$, and eigenvalues $\lambda_\phi^{\pm}$. Edge Majorana modes with $\phi=0,\pi$ exist in the parts of phase diagrams mentioned above. Below we explicitly write the wave function of the $0$ and $\pi$ Majorana modes. 

The eigenvalue of the transfer matrix with $|\lambda_\phi|<1$ is given by: 
\begin{equation}\label{eq:lambda}
    \lambda_{0,\pi} =
    \frac {[\cos(\pi J)+1][-\cos(\pi g)\pm1]}
    {\sin(\pi J)\sin(\pi g)}
    =
    \left\{
    \begin{split}
        & \frac{\tan \frac{\pi g}{2}}{\tan \frac{\pi J}{2}}
        & \quad \text{for } \phi=0
        \\
        & - \frac{1}{\tan \frac{\pi g}{2} \tan \frac {\pi J}{2}}
        & \quad \text{for } \phi=\pi
        \\
    \end{split}
    \right.
    .
\end{equation}

The wave function of the zero Majorana mode at the left edge of the chain reads: 
\begin{equation}\label{eq:pi_wf0}
\hat\chi_{\rm L}^0=C_0 \sum_j \lambda_0^{j-1} (\cos \frac {\pi g}{2} \, a_{2j-1} + \sin \frac{\pi g}{2} \, a_{2j} ), \;\;\; C_0=\sqrt{1-\lambda_0^2}.
\end{equation}

In terms of spin variables, this can be written as follows:
\begin{equation}\label{eq:spin_wf0}
\hat\chi_{\rm L}^0=C_0 \left[ \cos\frac {\pi g}{2} \hat{Z}_1+\sin \frac {\pi g}{2} \hat{Y}_1 + \lambda_0  \cos \frac {\pi g}{2} \hat{X}_1 \hat{Z}_2 + \lambda_0  \sin\frac {\pi g}{2} \hat{X}_1 \hat{Y}_2+...  \right]
\end{equation}

Similarly, we obtain the wave function of the $\pi$-Majorana mode at the left edge:
\begin{equation}\label{eq:pi_wf}
\hat\chi^\pi_{\rm L}=C_\pi \sum_j \lambda_\pi^{j-1} (\sin \frac{\pi g}{2} \, a_{2j-1} - \cos\frac{\pi g}{2} \, a_{2j} ), \;\;\; C_\pi=\sqrt{1-\lambda_\pi^2}.
\end{equation}

In the original spin variables, the $\pi$-Majorana operators read
\begin{equation}\label{eq:spin_wf}
\hat\chi^\pi_{\rm L}=C_\pi \left[ \sin\frac{\pi g}{2} \hat{Z}_1-\cos\frac{\pi g}{2}\hat{Y}_1 + \lambda_\pi \sin\frac{\pi g}{2} \hat{X}_1 \hat{Z}_2 -\lambda_\pi \cos\frac{\pi g}{2} \hat{X}_1 \hat{Y}_2+...  \right], 
\end{equation}
which corresponds to Eq.(3) of the main text. 

We ntoe that the Majorana wave function decays exponentially into the bulk, $\chi_\pi (j)\propto e^{-j/\xi_\pi}$, with localization length related to $\lambda_\pi$ as follows,
\begin{equation}\label{eq:loc_length}
\xi_\pi=-\frac{1}{\ln \lambda_\pi}
\end{equation}

The Majorana operator $\hat\chi_{\rm R}^\pi$ on the right edge can be obtained from Eq.(\ref{eq:pi_wf}) for $\hat\chi_{\rm L}^\pi$ by changing indices of the Majorana operators $a_{2j-1}, a_{2j}$ via $a_{2j-1}\to a_{2(L-j+1)}, a_{2j}\to a_{2(L-j)+1}$, where $L$ is the number of sites. In the spin language, this yields:
\begin{equation}\label{eq:chiR}
    \hat\chi_{\rm R}^\pi=C_\pi {\mathcal S} [\sin\frac{\pi g}2 \hat{Z}_L -\cos\frac{\pi g}2 \hat{Y}_L +\lambda_\pi \sin\frac{\pi g}2 \hat{X}_{L}\hat{Z}_{L-1}-\lambda_\pi \cos\frac{\pi g}2 \hat{X}_L \hat{Y}_{L-1}+... ],
\end{equation}
where ${\mathcal S}$ is the symmetry operation of flipping all spins, 
\begin{equation}
    {\mathcal S}=\prod_{i=1}^L X_i. 
\end{equation}
Since $[\hat{U}_{\rm F},{\mathcal S}]=0$, ${\mathcal S}\hat\chi_{\rm R}^\pi$ is also a $\pi$ Marjorana operator localized on the right edge (in contrast, $\hat\chi_{\rm R}^\pi$ has a Jordan-Wigner string). 

\subsection{Hybridization of edge modes in a finite chain and $\pi$-eigenstate pairing}

In this Subsection, we discuss the tunnel splitting of edge Majorana modes in finite chains, as well as its implications for the many-body quasienergy spectrum. 

The exponentially localized Majorana operators are eigenmodes of the Floquet operator $\hat{U}_{\rm F,0}$ in a semi-infinite chain. In a finite chain in the fermionic representation, left and right Majorana modes hybridize. The hybdridized modes 
$$d=(\hat\chi_{\rm L}^\pi + i \hat\chi_{\rm R}^\pi)/{2}, \;\; d^\dagger=(\hat\chi_{\rm L}^\pi - i \hat\chi_{\rm R}^\pi)/{2}, $$ 
have quasienergies $\pi\pm \Delta(L)/2$, which are tunnel-split, with 
\begin{equation}
    \Delta(L)\propto \exp(-L/\xi_\pi). 
\end{equation}
This has an interesting implication for the structure of many-body eigenstates~\cite{Mitra19}. A many-body eigenstate corresponds to a given occupation of non-interacting fermionic modes. Thus, two many-body eigenstates $|\theta\rangle$ and $|\tilde\theta\rangle$ that differ in the occupation of $d$-level (e.g. it is empty in $|\theta\rangle$, i.e. $d|\theta\rangle=0$ and occupied in $|\tilde\theta\rangle$, $d^\dagger|\tilde\theta\rangle$=0), will have a quasienergy difference 
\begin{equation}
    \delta\theta=\pi+\Delta(L). 
\end{equation}
Since $\Delta(L)\to 0$ as $L\to \infty$, in the limit of a very long chain, eigenstates become ``$\pi$-paired" -- that is, each eigenstate has a partner state with quasienergy shifted by $\pi$. 

Importantly, one can obtain the partner state from $|\theta\rangle$ by acting with either of the Majorana operators approximately localized on either edge, since e.g. $\hat\chi_{\rm L}^\pi=d+d^\dagger$. Thus, $\hat\chi_{\rm L}^\pi|\theta\rangle=d^\dagger|\theta\rangle=|\tilde\theta\rangle$. Note that this relation holds equally for the fermionic and spin representation of the model. 


\section{Robustness of edge modes with respect to perturbations}

While conserved Majorana operators have been found exactly in a solvable model that maps to free fermions, an important question is to understand their fate when perturbations that break integrability are introduced. A related question is regarding the effect of such perturbations on the $\pi$ eigenstate pairing. We will consider the effect of perturbations of the Floquet drive which are periodic in time. We emphasize that Majorana operators are also expected to be robust with respect to slowly varying (on the scale of one Floquet driving period) perturbations, but not to errors happening on the time scale of one period, or faster. 

Below we will first describe general rigorous results giving bounds on the stability of edge Majorana operators with respect to non-integrable perturbations of the Floquet operator, and then provide numerical results illustrating the general bounds. The general physical mechanism is that of prethermalization~\cite{PRBPrethermal2017,MoriPRL16_RigorousBoundHeating,PrethermalRigorous}. We note that, following the work of Fendley~\cite{Fendley2016}, Else et al.~\cite{FendleyPRXPreth} showed that, provided certain conditions are met, prethermalization mechanism can protect edge modes in non-integrable Hamiltonian models, such as transverse-field Ising model with small perturbations, turning them into (almost) strong zero modes -- that is, modes with exponentially long lifetime. Closer to the context of interest to us, recently Ref.~\cite{Mitra19} numerically studied robustness of edge modes in a kicked Ising chain with a $Z_2$ spin-flip-symmetric perturbation. Below, we will be interested in generic perturbations, including those which break $Z_2$ symmetry.  

\subsection{Prethermalization}

We will use ideas of prethermalization in two ways -- first, to analyze the properties of the non-integrable Floquet operators, and second, to demonstrate the robustness of edge operators using an argument proposed in~\cite{FendleyPRXPreth}. First, we will need to recall the basic result by Else et al.~\cite{ElsePrethermalTimeCrystalPRX}, generalizing the results in Ref.~\cite{PRBPrethermal2017,PrethermalRigorous}. We quote this result somewhat colloquially: 

\vspace{0.1cm}

\noindent Suppose we have a Floquet operator 
$$
\hat{U}_{\rm F}=\hat{\mathcal{G}} {\mathcal{T}}\exp -i\int_0^T \hat{W}(t) dt,
$$
where ${\mathcal{T}}$ denotes time-ordering, $\hat{\mathcal{G}}^K=1$ for some integer $K>1$, and $\hat{W}(t)$ is a sum of local terms with a typical local energy scale $\lambda$. We assume that the latter scale is small, such that $\lambda T\ll 1$. Then, after dressing $\hat{U}_{\rm F}$ with a quasi-local unitary transformation $\mathcal{U}$, one can approximate it as follows:
\begin{equation}\label{eq:Fpreth}
{\mathcal U} \hat{U}_{\rm F} {\mathcal U^\dagger} \approx \hat{\mathcal{G}} e^{-i\hat{D}T}, 
\end{equation}
where $[\hat D,\hat{\mathcal{G}}]=0$. The error of this approximation is exponentially small in $1/\lambda T$. Thus, its effects start to affect physical observables only at parametrically long times of the order
$$
\tau\sim e^{C/(\lambda T)},
$$
where $C$ is a constant of order one. 

\vspace{0.1cm}

Importantly, this result guarantees that one can find $\hat{D}$ which respects the symmetry defined by $\hat{\mathcal{G}}$. Thus, even if $\hat W(t)$ breaks the symmetry, it will be effectively approximately restored. 

\subsection{Application to perturbed kicked Ising chain}

To apply this result to our setting, let us consider a perturbed drive:
\begin{equation}\label{eq:UFS}
    \hat{U}_{\rm F}= e^{-\frac{i\pi J}{2}\sum_{i=1}^{L-1} \hat Z_i \hat Z_{i+1}} e^{-\frac{i\pi g}{2} \sum_{i=1}^L \hat X_i} e^{-i\hat V}, 
\end{equation}
where $\hat V$ is a perturbation that is a sum of local terms $\hat V=\sum_{i=1}^L V_i$, which have a norm $||V_i||\leq |V|$. 

An example of a perturbation $\hat V$ of the integrable Floquet operator (\ref{eq:KIC}) relevant for our experiment is a sum of local $z$-fields, 
\begin{equation}\label{eq:Vrandomh}
\hat V=\frac{1}{2}\sum_{i=1}^L h_i \hat Z_i. 
\end{equation}
Here $h_i$ in general depends on the spin/qubit number. Note that this perturbation breaks the $\mathbf{Z}_2$ spin-flip symmetry of the model. 

We will show that the prethermalization results guarantee the robustness of the edge Majorana operators for a general perturbation $\hat{V}$ in two different limits: {\it (i)} $ \max (|\epsilon|, |V|) \ll|J|\ll 1$, where $\epsilon=1-g$, and {\it (ii)} $J=1/2, \max(|\epsilon||,|V|)\ll 1$. 

Importantly, for the experimentally relevant perturbation (\ref{eq:Vrandomh}), the robustness of the Majorana modes can be proven under a weaker condition, $|\epsilon|\ll 1$ -- that is, $|h_i|$ may be of order one. Below, we will first analyze the case of a generic perturbation, followed by a discussion of the perturbation in Eq.~(\ref{eq:Vrandomh}). 

\noindent {\it (i).} In the limit 
$$
g=1-\epsilon, \;\;  |\epsilon|, |V| \ll |J|\ll 1, 
$$
the conditions of the theorem are satisfied, with the parameter given by $\lambda=\max (|\epsilon|, |V|)$. By performing a (perturbatively constructed) quasilocal unitary transformation, we can bring the Floquet operator to the form (\ref{eq:Fpreth}) with 
$$
\hat{\mathcal{G}}=\prod_i \hat X_i, \;\;\; 
\hat D\approx \frac{\pi J}{2}\sum_i \hat Z_i \hat Z_{i+1}+ \hat D', \;\;\; \hat D'=O(|\epsilon|,|h|).
$$
In the above formula, $\approx$ sign indicates an error that is exponentially small (in this case, in $1/\max(|\epsilon|,|V|)$). 

We next apply the results of Ref.~\cite{PRBPrethermal2017} to this operator. These results indicate that the number of {\it domain walls}, $N_{dw}=\sum \hat Z_i \hat Z_{i+1}$ is approximately conserved, up to a parametrically long time. Then, via an argument analogous to that in Ref.~\cite{FendleyPRXPreth}, this approximate conservation law implies the existence of an almost conserved edge operator $\hat\chi^\pi$, which anti-commutes with the Floquet operator. This Majorana operator has a lifetime bounded from below by
$$
\tau' \sim e^{CJ/f(\epsilon,|V|)},
$$
where $f(\epsilon,|V|)\to 0$ as $\epsilon\to 0$ or $|V|\to 0$. We therefore expect the Majorana operator to be most robust with respect to generic integrability-breaking fields when $\epsilon$ is small, that is, when parameter $g$ is close to $1$. 

\noindent {\it (ii).} The second limit, which is relevant for the experimental measurements, is specified by the conditions 
$$
J=1/2 \;\;\; \max(|\epsilon|,|V|)\ll 1.
$$
In this case, we choose
$$
\hat{\mathcal{G}}=e^{-i\pi/4\sum_{i=1}^{L-1} \hat{Z}_i \hat{Z}_{i+1}} \prod_{i=1}^L \hat{X}_i.
$$
Using the fact that $\prod_i \hat X_i$ commutes with operator $\sum \hat Z_i \hat Z_{i+1}$, we obtain that $\hat{\mathcal{G}}^K=1$ with $K=8$. 

Following similar steps as in the regime {\it (i)}, we can show that $e^{-i\pi/4\sum_{i=1}^{L-1}\hat Z_i \hat Z_{i+1}}$ is an approximate prethermal conservation law. This is once again sufficient to guarantee the robustness of prethermal edge operators. 

Finally, we discuss the case of random $z$-fields, see Eq.(\ref{eq:Vrandomh}). First, we perform a unitary transformation of the Floquet operator, with a unitary $\hat{\mathcal{U}}_1=e^{-\frac{i}{4}\sum_{i=1}^L h_i \hat Z_i}$: 
\begin{equation}\label{eq:UF'}
\hat{U}_F'=\hat{\mathcal{U}}_1 \hat{U}_F \hat{\mathcal{U}}_1^\dagger= e^{-\frac{i}{4}\sum_{i=1}^L h_i \hat Z_i} 
e^{-\frac{i\pi J}{2}\sum_{i=1}^{L-1} \hat Z_i \hat Z_{i+1}} e^{-\frac{i\pi g}{2} \sum_{i=1}^L \hat X_i} e^{-\frac{i}{4}\sum_{i=1}^L h_i \hat Z_i}. 
\end{equation}
As we now argue, this unitary transformation brings the Floquet operator to the form (\ref{eq:UFS}), with $|V|\sim O(|\epsilon|)$. This can be seen by rewriting the last two terms in the above equation using an identity for the Pauli operators:  
$$
e^{-\frac{i\pi g}{2}  \hat X_i} e^{-\frac{i}{4} h_i \hat Z_i}=e^{\frac{i\pi \epsilon}{2} \hat X_i} e^{-\frac{i\pi}{2} \hat X_i} e^{-\frac{i}{4} h_i \hat Z_i}=e^{\frac{i}{4} h_i \hat Z_i} e^{\frac{i\pi\epsilon}{2} (\cos(h_i/2)\hat X_i+\sin(h_i/2) \hat Y_i)}e^{-\frac{i\pi}{2} \hat X_i}. 
$$
This brings Eq.(\ref{eq:UF'}) to the following form: 
\begin{equation}\label{eq:UF'2}
\hat{U}_F'=e^{-\frac{i\pi J}{2}\sum_{i=1}^{L-1} \hat Z_i \hat Z_{i+1}} e^{-\frac{i\pi}{2} \sum_{i=1}^L\hat X_i}
e^{\frac{i\pi\epsilon}{2} \sum_{i=1}^L (\cos(h_i/2)\hat X_i+\sin(h_i/2) \hat Y_i)},
\end{equation}
which is identical to Eq.(\ref{eq:UFS}), with $|V|\sim O(|\epsilon|)$, even when $|h_i|\sim 1$. Then, following the line of argument described above for a generic perturbation, we conclude that the Majorana edge operator is robust up to an exponentially long time. 

\begin{figure}[t]
  \centering
  \includegraphics[width=1\columnwidth]{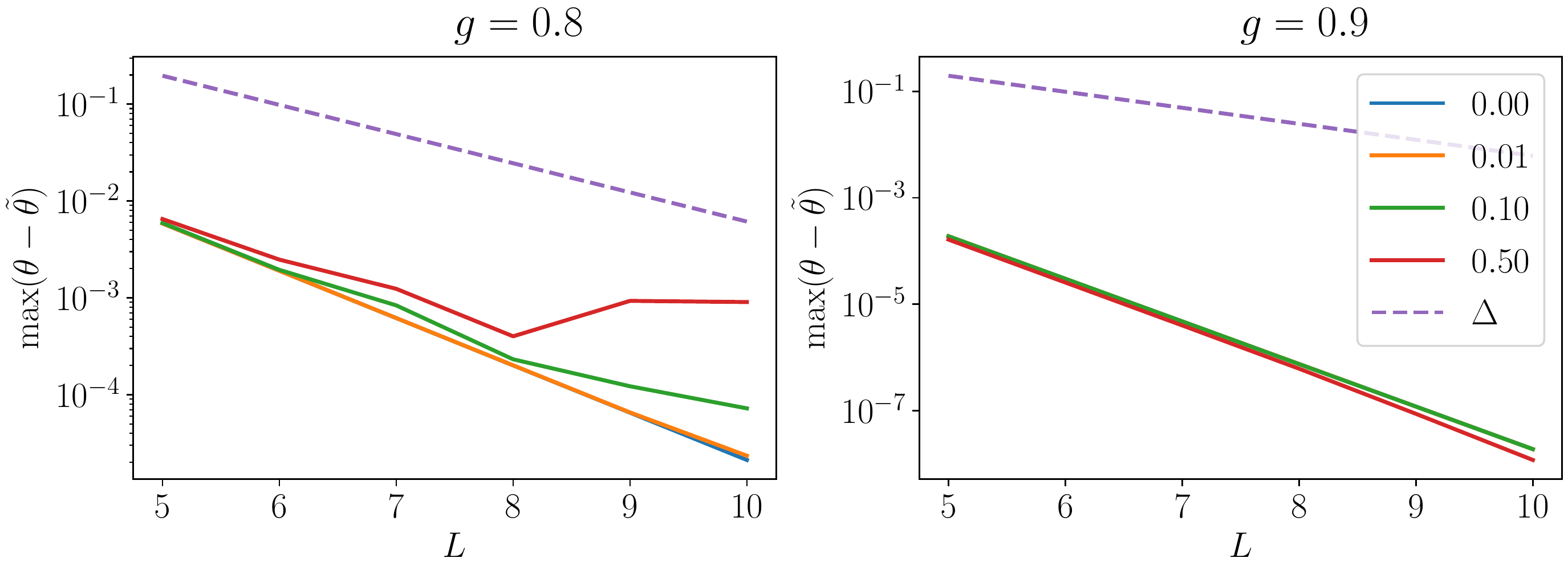} 
 \caption{%
Maximum quasienergy difference between an eigenstate and its $\pi$-partner, as a function of system size $L$ and for varying strength of the integrability-breaking $Z$-field, shown in the inset. For $g=0.9$, the integrability breaking has virtually no effect on the $\pi$-pairing, while for $g=0.8$ the $\pi$-pairing is robust up to a certain threshold value of the $Z$-field. Here $\Delta$ is the average quasienergy level spacing. 
 }
 \label{fig_splitting}
\end{figure}

To summarize this subsection, we argued that, irrespective of the form of a time-periodic perturbation, the edge Majorana operator generally becomes approximately conserved, with a decay rate that is non-analytic in the perturbation strength. For the case of random $z$-fields, this robustness extends to the regime when $h_i$ become of order one. 

\subsection{Correction to the Majorana operator due to weak $z$-fields}\label{sec:perturb}

We argued above that for an arbitrary non-integrable perturbation, Majorana edge operator, dressed by a quasilocal unitary transformation ${\mathcal U}$, remains long-lived up to an exponentially long time. The unitary transformation modifies the structure of the edge operator, giving rise to new long-lived correlators, as demonstrated in the main text. Below we consider a Floquet operator that involves $Z$-fields,
\begin{equation}
\hat{U}_{\rm F}= \hat{U}_{F,0} e^{-\frac{i}{2}\sum_{i=1}^L h_i \hat Z_i} \equiv \hat{U}_{F,0} \hat{K}, 
\end{equation}
where $\hat{U}_{\rm F,0}$ is an unperturbed Floquet operator. Our aim is to compute the perturbative (in $h_i$) correction to the $\pi$-Majorana operator. We look for a corrected Majorana operator at the left edge in the following form (for simplicity dropping subscript $L$):
$$
\tilde\chi^\pi=\hat\chi^\pi+\delta \hat\chi, 
$$
choosing $\delta\hat\chi$ to satisfy relation
\begin{equation}
\hat K \hat{U}_{\rm F,0}^{-1} (\hat\chi^\pi+\delta \hat\chi) \hat{U}_{\rm F,0} \hat K =-(\hat\chi^\pi+\delta\hat\chi).
\end{equation}
Using $\hat{U}_{\rm F,0}^{-1} \hat\chi^\pi \hat{U}_{\rm F,0}=-\hat\chi^\pi$, we obtain
$$
\hat K(-\hat\chi^\pi +\hat{U}_{\rm F,0}^{-1} \delta\hat\chi \hat{U}_{\rm F,0}) \hat K^{-1} =-(\hat\chi^\pi+\delta\hat\chi).  
$$
Further, assuming $\delta\hat\chi =O(|h|)$, $|h|\ll 1$, we can approximate
$$
\hat K \hat\chi^\pi \hat K^{-1}\approx \hat\chi^\pi -\frac i2 [\sum h_i \hat Z_i, \hat\chi^\pi]. 
$$ 
Therefore, 
$$
\hat{U}_{\rm F,0}^{-1} \delta\hat\chi \hat{U}_{\rm F,0} +\delta\hat\chi = -\frac i2 [\sum h_i \hat Z_i, \hat\chi^\pi]\approx -\cos \frac{\pi g}{2} h_1 [-\hat X_1+\hat Y_1 \hat Z_2], 
$$
where we assumed $\epsilon\equiv 1-g \ll 1$, and took $J=\pi/4$, as in our experiment. The latter equation holds to the first order in $|\epsilon|$. 

Then, we can check that the 1st correction to the $\pi$-mode, which satisfies the above equation, is given by: 
\begin{equation}
\delta \chi=-\cos \frac{\pi g}{2} h_1 \hat Y_1 \hat Z_2. 
\end{equation}
Two comments are in order: first, we see that in the lowest order in $|h|, |\epsilon|$, only the field $h_1$ at the site 1 contributes to the correction; second, the correction will average to zero if averaging over $h_1$ is performed, as is done in the some of the experiments described above. 

The implication of this result is that compared to the integrable case, a new long-lived correlator, involving $\hat Y_1(t)\hat Z_2(t)$, appears. This provides an explanation for the experimental findings discussed in the main text.

\subsection{$\pi$-pairing in a non-integrable model}

So far, we discussed the robustness of edge operators in terms of their lifetimes. An interesting question concerns the $\pi$-pairing of eigenstates in the presence of non-integrable perturbations. The arguments above show that dressed Majorana operator $\tilde\chi^\pi$ anticommutes with $\hat{U}_{\rm F}$, up to an error term that is exponentially small in the perturbation strength. Therefore, we expect that in a finite-size system $\pi$-pairing will persist up to some maximum system size $L_*$, as long as the level spacing is large compared to the exponentially small error term $\tilde\chi^\pi \hat{U}_{\rm F}+\hat{U}_{\rm F} \tilde\chi^\pi$. 

For illustration purposes, we numerically computed the maximum of $\Delta\theta_{\rm max}= |\theta-\tilde\theta|$ over all eigenstates. Here $\theta$ and $\tilde\theta$ are the quasienergies of the two eigenstates that are $\pi$-partners. The evolution of this quantity with constant $Z$-field (we put $h_i=2h$ for all spins, note the factor of $2$) is illustrated in Fig.~\ref{fig_splitting}. In agreement with our expectation, for $g=0.9$ (where $\epsilon\ll 1$), the $\pi$-pairing is insensitive even to relatively strong integrability breaking perturbation. At smaller $g$, we observe that for a fixed $h$ $\pi$-pairing starts to disappear above some system size $L_*(h)$, which is a decreasing function of $h$. We have also checked that the eigenstate pairing exhibits similar robustness with respect to random $Z$-fields (not shown).

\section{Reconstructing single-particle spectrum}

In this Appendix, we show that for an integrable Floquet drive, the Fourier transform of $\langle \hat Z_1(t) \hat Z_1 (0) \rangle$ correlator allows us to reconstruct the single-particle excitation spectrum. We consider an initial state $|\psi(0)\rangle$ which is a bit-string state, with the edge spin being in the $Z_1=+1$ state:  
\begin{equation}\label{eq:Z1}
\hat Z_1|\psi(0)\rangle=|\psi(0)\rangle. 
\end{equation}
Then 
\begin{equation}\label{eq:CZZ_spectrum}
C_{ZZ}(t)=\langle \psi(0)| \hat Z_1(t)\hat Z_1(0) |\psi(0)\rangle =\langle \psi(0)| \hat{U}_{\rm F,0}^{-t} \hat Z_1 \hat{U}_{\rm F,0}^t |\psi(0)\rangle,     
\end{equation}
Operator $\hat Z_1$ can be expressed via $c,c^\dagger$ fermionic operators, 
$$
\hat Z_1=i(c_1^\dagger-c_1). 
$$
Further, we rewrite this expression via the Bogoliubov eigenmodes $\gamma_\alpha, \gamma^\dagger_\alpha$ of the Floquet operator (\ref{eq:Ffermions}), defined by relations
$$
\hat{U}_{\rm F,0}^{-1} \gamma_\alpha \hat{U}_{\rm F,0}=e^{-i\phi_\alpha} \gamma_\alpha, \;\; \hat{U}_{\rm F,0}^{-1} \gamma_\alpha^\dagger \hat{U}_{\rm F,0}=e^{i\phi_\alpha} \gamma_\alpha^\dagger, 
$$
which gives
$$
\hat Z_1=\sum_\alpha u_\alpha \gamma_\alpha +u_\alpha^* \gamma_\alpha^\dagger, 
$$
where $\sum_\alpha |u_\alpha|^2=1$, and, as follows from (\ref{eq:Z1}), $\sum_\alpha\langle\psi(0)| u_\alpha \gamma_\alpha+u_\alpha^*\gamma_\alpha^\dagger  |\psi(0)\rangle=1$. 
Next, we observe that 
$$
C_{ZZ}(t)=\sum_\alpha u_\alpha e^{-i\phi_\alpha t} \langle\psi(0)|\gamma_\alpha |\psi(0) \rangle +    
u_\alpha^* e^{i\phi_\alpha t} \langle\psi(0)|\gamma_\alpha^\dagger |\psi(0) \rangle.
$$
Thus, the Fourier transformation of $C_{ZZ}$ contains frequencies identical to the (single-particle) quasienergies. The height of each Fourier peak depends both on the initial state and on the coefficients in the expansion of original fermionic operators $c_1,c_1^\dagger$ in terms of the Bogoliubov eigenmodes. Nevertheless, as shown in the main text, this allows us to extract the single-particle quasienergy spectrum and detect gap closing at $g=J=1/2$. 

\section{Extracting the Majorana operator}

In this Section, we describe the method of extracting a local integral of motion, based on measuring Pauli strings. A local integral of motion (LIOM) can be expanded as follows:
\begin{equation}{\label{eq:structure}}
\hat\chi=\sum \alpha_{\kappa } \hat{O}_{\kappa },  \,\, \sum \alpha_{\kappa}^2=1. 
\end{equation}
where $\hat{O}_{\kappa}$ are Pauli strings (with $\hat O_1$ being the most local term). 

Consider a setup where our initial state $|\psi(0)\rangle$ is such that
\begin{equation}
\langle \psi(0)| \hat O_1|\psi(0) \rangle=1, 
\end{equation}
and
\begin{equation}
\langle \psi (0) | \hat O_\kappa|\psi (0) \rangle=0, \,\, \kappa \geq 2.
\end{equation}
Note that our setup satisfies this requirement, since there $\hat O_1=\hat Z_1$ and we consider initial bit strings. 

Then, the conserved operator will obey 
$$
\langle \hat\chi(t) \rangle = \langle  \hat\chi(0) \rangle =\alpha_1. 
$$
Next, we note that 
$$
\hat O_\kappa = \alpha_\kappa \hat\chi + \sum_j \beta_j \hat\eta_j, 
$$
where $\hat\eta_j$ is an operator orthogonal to $\hat\chi$, $\text{Tr} (\hat\chi \hat\eta_j)=0$. 
If $\hat\eta_j$ operators spread/thermalize then their expectation value is zero at sufficiently long times; then, the saturated value of $\hat O_\kappa$ will be:
$$
\langle \hat O_\kappa \rangle _{\infty}=\alpha_\kappa \langle  \hat\chi(0)\rangle =\alpha_\kappa \alpha_1. 
$$
Therefore, under this assumption (that we have just one local integral of motion and we waited long enough and operator values reached saturation), the saturation values alone give us all the information about the conserved operator. In particular, for the $\pi$ edge Majorana mode, we expect the following values: 
$$
\alpha_{Z_1}=C_\pi \sin \frac{\pi g}{2}, \,\,  \alpha_{Y_1}=-C_\pi \cos \frac{\pi g}{2}, \,\,   \alpha_{X_1 Z_2}= C_\pi \lambda_\pi \sin \frac{\pi g}{2}  , \,\,  \alpha_{X_1 Y_2}=-C_\pi \lambda_\pi \cos \frac{\pi g}{2}, \,\, 
 \alpha_{X_1 X_2 Z_3}=C_\pi \lambda_\pi^2 \sin\frac{\pi g}{2}, \,\, 
$$
where $\lambda_\pi$ is defined in Eq.~(\ref{eq:lambda}). As discussed below, weak dissipation gives rise to a slow temporal decay of expectation values $\langle \psi(0)| \hat O_\kappa |\psi(t) \rangle$, with identical rates. Nevertheless, the ratios of these expectation values allow us to determine coefficients $\alpha_\kappa$.

\begin{figure}[t]
  \centering
  \includegraphics[width=0.5\columnwidth]{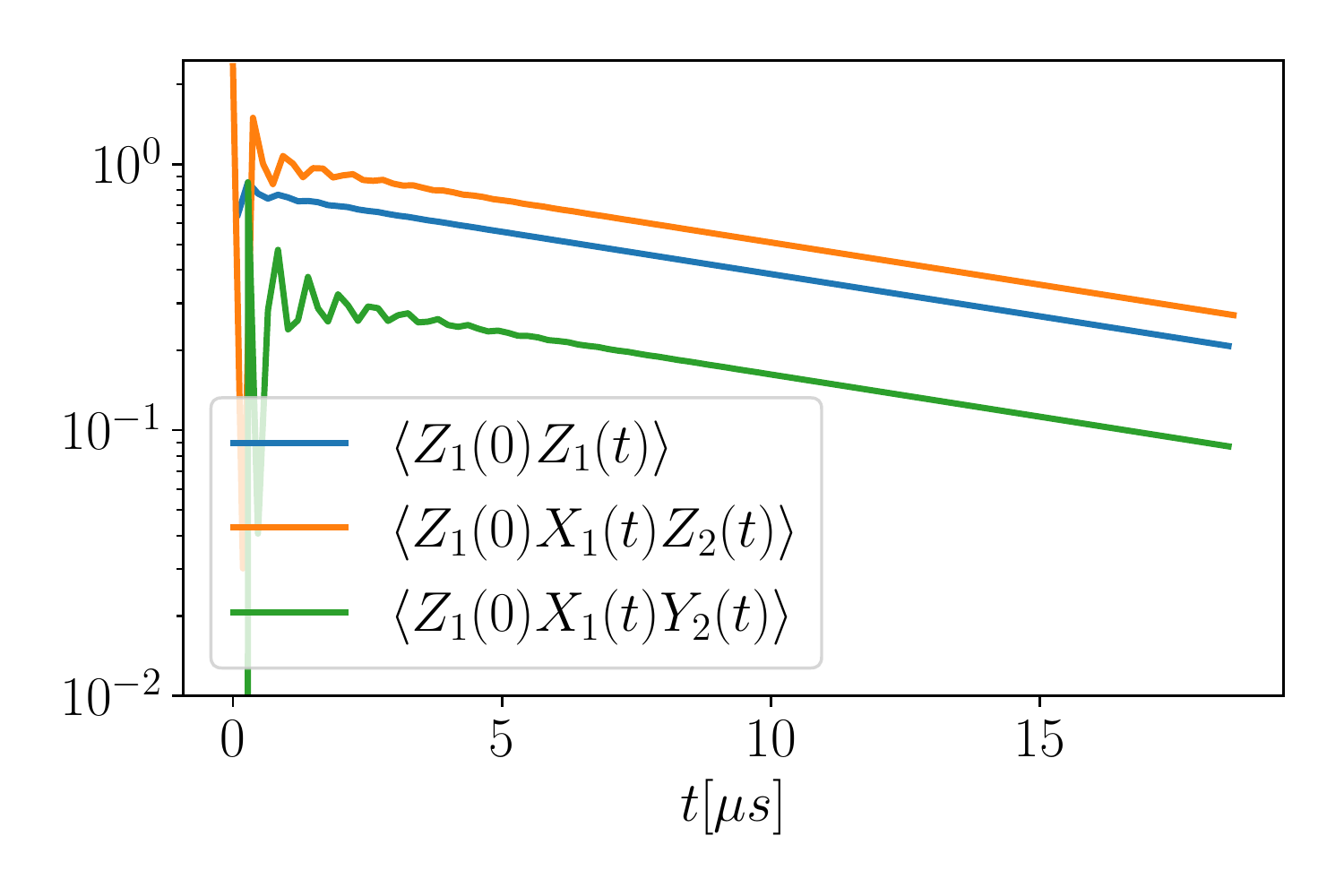} 
 \caption{%
 Time-dependent expectation values of operators that enter the Majorana operator (\ref{eq:spin_wf}) in the presence of dissipation. All correlators exhibit short-time transient, follows by an exponential decay with identical rates. This is in agreement with experimental observations and theoretical analysis in Sec.~\ref{app:dissipation}. The calculations were performed using influence matrix method, described in Sec.~\ref{app:numerics} in the limit of an infinitely long chain. Parameters were chosen as follows: $J=1/2, g=0.8, \gamma_\phi=0.01, \gamma_d=0.0046$.
 }
 \label{fig:corr_decay}
\end{figure}

\section{Effect of decay and dephasing} \label{app:dissipation}

In this Section, we discuss the effect of single-qubit decay and dephasing on the temporal correlations of operators entering the Majorana operators (\ref{eq:spin_wf0},\ref{eq:spin_wf}). Under an approximation that local operators which do not enter in this expansion spread quickly, we argue that correlators $\langle O_i(t) O_j (0) \rangle$ which we measure experimentally, decay with an identical rate. This is a key theory result which allows us to reconstruct the Majorana operator despite dissipation. 

Below we will study Hamiltonian case, considering a semi-infinite system and one (e.g. left) edge operator $[\chi,H]=0$. Extension of the analysis to the Floquet case is straightforward. 

We start with the Lindblad equation desribing evolution of the system's density matrix $\rho$:
\begin{equation}\label{eq:lindblad}
    \frac{d\rho}{dt}={\mathcal L} [\rho]=-i[H,\rho]+\sum_\nu L_\nu \rho L_\nu^\dagger -\frac{1}{2} \{ L_\nu^\dagger L_\nu, \rho \}, 
\end{equation}
where $L_\nu$ are the jump operators. 
For our purposes, it is more convenient to consider the operator evolution in Heisenberg representation. An operator $W$ evolves according to the following equation: 
\begin{equation}\label{eq:lindblad_op}
\frac{dW}{dt}={\mathcal L}^* [W(t)]=i[H,W(t)]+\sum_\nu L_\nu^\dagger W(t) L_\nu -\frac{1}{2} \{ L_\nu^\dagger L_\nu, W(t) \}. 
\end{equation}
We will consider jump operators which describe dephasing and decay on each site $j$: 
\begin{equation}\label{eq:jumps}
    L_{j\phi}=\sqrt{\frac{\gamma_{j \phi}}{2}} Z_j, \;\; L_{jd}=\sqrt{\gamma_{jd}} \sigma^-. 
\end{equation}
Throughout our analysis, we will assume that dephasing and decay are relatively weak, such that the corresponding rates are slow on the scale of internal unitary dynamics of the system. It will be also convenient to sometimes separate the conjugate Lindbladian ${\mathcal{L}}^*$ into the Hamiltonian part and dissipative part: 
\begin{equation}\label{eq:Ldiss}
{\mathcal{L}}^*[\cdot]=i[H,\cdot]+{\mathcal L}_{\rm diss}^* [\cdot].
\end{equation}

To analyze the decay of the conserved operator $\chi$, let us consider an operator basis that includes $\chi$ and operators $B_\mu$ that are orthogonal to it (with respect to the trace inner product).  An arbitrary operator $W$ can be expanded as a linear combination of $\chi, B_\mu$:
\begin{equation}\label{eq:W}
    W=C_\chi \chi+\sum_\mu C_\mu B_\mu. 
\end{equation}
Plugging this into Eq.(\ref{eq:lindblad_op}), we will obtain a system of linear equations for $C$ coefficients. 

We will need to first analyze the action of ${\mathcal L}^*$ on $\chi$. To that end, let us write down the action of ${\mathcal L}^*_{\rm diss}$ on the Pauli operators:
\begin{equation}\label{eq:lindblad_pauli}
 {\mathcal L}^*_{\rm diss} X_j=-\Gamma_{j}X_j, \;\;   {\mathcal L}^*_{\rm diss} Y_j=-\Gamma_{j}Y_j, \;\; {\mathcal L}^*_{\rm diss} Z_j=-\gamma_{jd}[1+Z_j],
\end{equation}
where we defined 
\begin{equation}\label{eq:G1}
\Gamma_j=\gamma_{j\phi}+\frac{\gamma_{jd}}2. 
\end{equation}
From these equations, we can write the action of ${\mathcal L}^*_{\rm diss}$ on the Pauli strings that enter the expansion of $\chi$ (see Eq.~\ref{eq:spin_wf0}): 
\begin{equation}\label{eq:lindblad_strings}
     {\mathcal L}^*_{\rm diss} X_1...X_{k-1}Z_k=-\left( \sum_{i=1}^{k-1} \Gamma_i +\gamma_{kd} \right) X_1...X_{k-1}[1+Z_k], \;\; 
     {\mathcal L}^*_{\rm diss} X_1...X_{k-1}Y_k=-\left( \sum_{i=1}^{k} \Gamma_i \right) X_1...X_{k-1}Y_k.  
\end{equation}
Let us denote the decay rate of a given Pauli string $O_{\{\kappa\}}$ by $\Gamma_{\{\kappa \}}$. Then, we can express the action of the dissipative part of the Lindbladian on $\chi$ as follows:
\begin{equation}\label{eq:l_chi}
    {\mathcal L}^*_{\rm diss} \chi=-\Gamma_{\rm eff} \chi+\sum_\mu c_\mu B_\mu, \;\; \Gamma_{\rm eff}=\sum_{\{\kappa\}} \alpha_{\{\kappa \}}^2 \Gamma_{\{\kappa\}}. 
\end{equation}
Note that $c_\mu=O(\gamma)$, and moreover the weight $B_\mu$ in the above formula decays exponentially with the support of $B_\mu$. 

Since $\chi$ is a conserved operator, $[H,\chi]=0$, and therefore the action of ${\mathcal L}^*$ on $\chi$ is the same as that of ${\mathcal L}^*_{\rm diss}$. Next, let us write the action of ${\mathcal L}^*$ on $B_\mu$. We first observe that $i[H,B_\mu]=i\sum_\nu A_{\mu\nu} B_\nu$, that is, $[H,B_\mu]$ is orthogonal to $\chi$ operator (this can be verified using the relation $[H,\chi]=0$). Therefore, 
$$
{\mathcal L}^* B_\mu=\sum_\nu A_{\mu\nu} B_\nu + b_\mu \chi, 
$$
with $b_\mu \chi$ term originating from the action of the dissipative part of the Lindbladian on $B_\mu$, and therefore $b_\mu=O(\gamma)$. 

From the above equations, we can obtain the system of linear equations for the evolution of a general operator $W$ in Eq.~(\ref{eq:W}):
\begin{equation}\label{eq:lindblad_matrix}
\left( \begin{array}{c}
    \dot{C}_\chi   \\
    \dot{C}_\mu    
 \end{array} \right)
 =
 \left( \begin{array}{cc}
-\Gamma_{\rm eff} & b_\nu \\  
c_\mu & A_{\mu\nu}  
 \end{array} \right) 
  \left( \begin{array}{c}
C_\chi \\   
C_\nu 
 \end{array} \right), 
\end{equation}
where we used a compact notation to label the matrix entries $A_{\mu\nu}, b_\mu, c_\nu$.

Next, we make an assumption that operators $B_\mu$ for which $b_\mu,c_\mu$ are sizeable, spread and or/dissipate quickly compared to the rate $\Gamma_{\rm eff}$. For the case when dynamics is chaotic it is expected that local operators rapidly evolve into complex superpositions of Pauli strings, which decay quickly due to dissipation. For the integrable case, there is a set of local integrals of motion, and some $B_\mu$ operators have overlap with those integrals of motion; however, we note that the terms entering integrals of motion carry Jordan-Wigner strings, which makes them more susceptible to dissipation, compared to $\chi$. Thus, operator $\chi$ is an approximate eigenmode of the Lindbladian with decay rate $\Gamma_{\rm eff}$. The corrections to this eigenvalue arise due to mixing with other modes, and are expected to be of the order $\gamma^2$. We leave a rigorous analysis of this phenomenon for a future study, but provide a numerical computation below. 

Thus, $\chi$ is a long-lived operator, since the decay rate $\Gamma_{\rm eff}$ is in general of the same order of magnitude as single-qubit decay/dephasing rates. Other local operators spread and decay much faster. A direct implication of this fact is that the Pauli strings entering the expansion of $\chi$, will exhibit {\it identical} long lifetimes, which are greatly exceed those of other Pauli strings. To confirm this expectation, we extended the influence matrix method for quantum many-body dynamics to incorporate dissipation~\cite{LerosePRX,SonnerAoP}, and applied it to compute the behavior of muilti-spin observables in the Majorana operator expansion. The result is illustrated in Fig.~\ref{fig:corr_decay}, and the details of the numerical procedure can be found in Sec.~\ref{app:numerics}.

Finally, we provide an explicit expression of $\Gamma_{\rm eff}$ for the $\pi$ Majorana operator: 
\begin{equation}\label{equation:Gamma_eff_pi}
    \Gamma_{\rm eff}=C_\pi^2\sin^2 \frac{\pi g}{2}\sum_{k=1}^\infty \lambda_\pi^{2(k-1)} \left(\sum_{i=1}^{k-1} \Gamma_i +\gamma_{kd}\right) +C_\pi^2\cos^2 \frac{\pi g}{2}\sum_{k=1}^\infty \lambda_\pi^{2(k-1)} \sum_{i=1}^k \Gamma_i
\end{equation}

\begin{figure}[t]
  \centering
  \includegraphics[width=0.5\columnwidth]{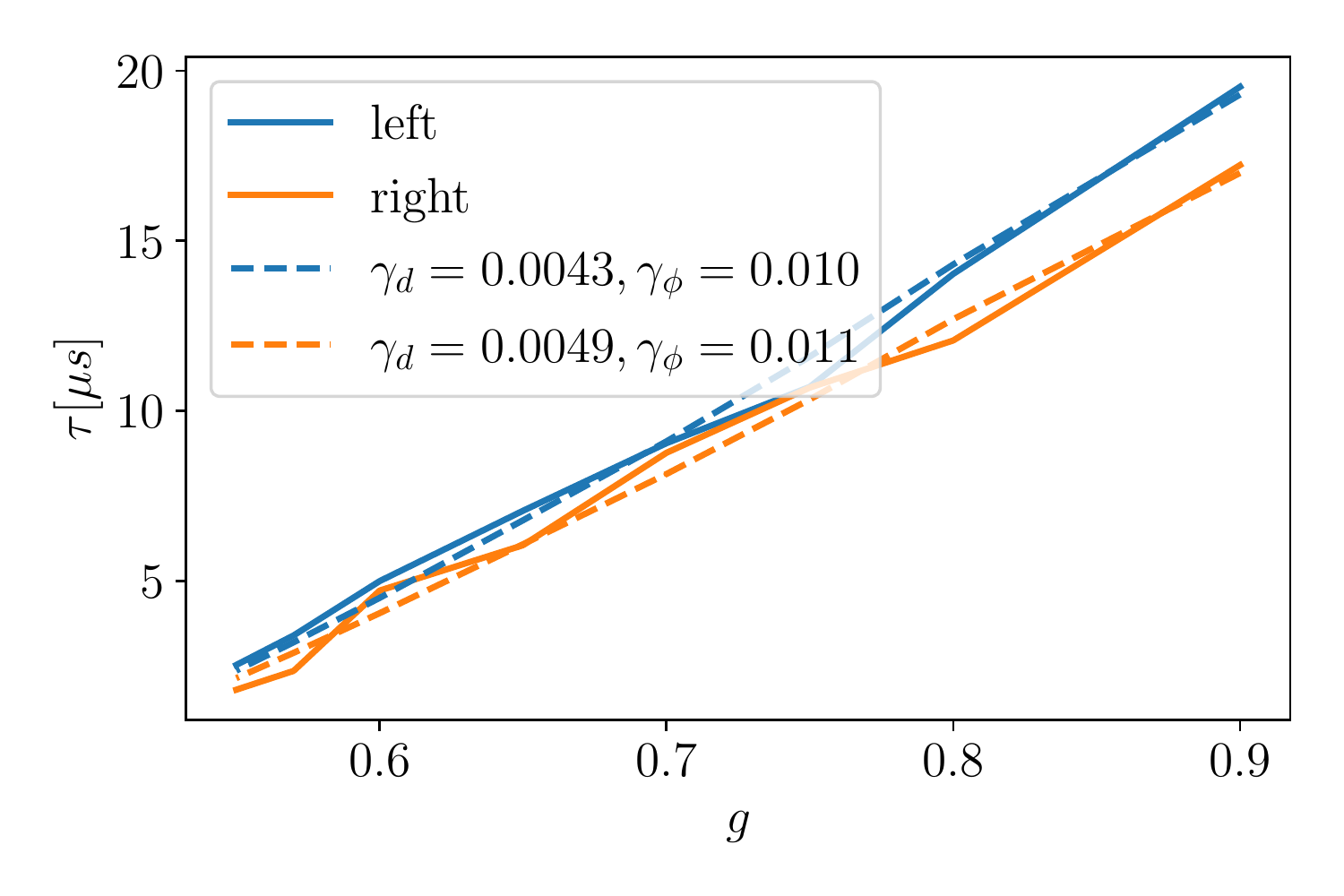} 
 \caption{%
Theoretical fit of the experimental data for the decay time of the Majorana operator, at fixed $J=1/2$, as a function of $g$. 
The decay and dephasing parameters $\gamma_\phi,\gamma_d$ were assumed to be uniform, and chosen to achieve the best fit according to least-square criterion. 
 }
 \label{fig:fit_rates}
\end{figure}

\begin{figure}[t]
  \centering
  \includegraphics[width=0.5\columnwidth]{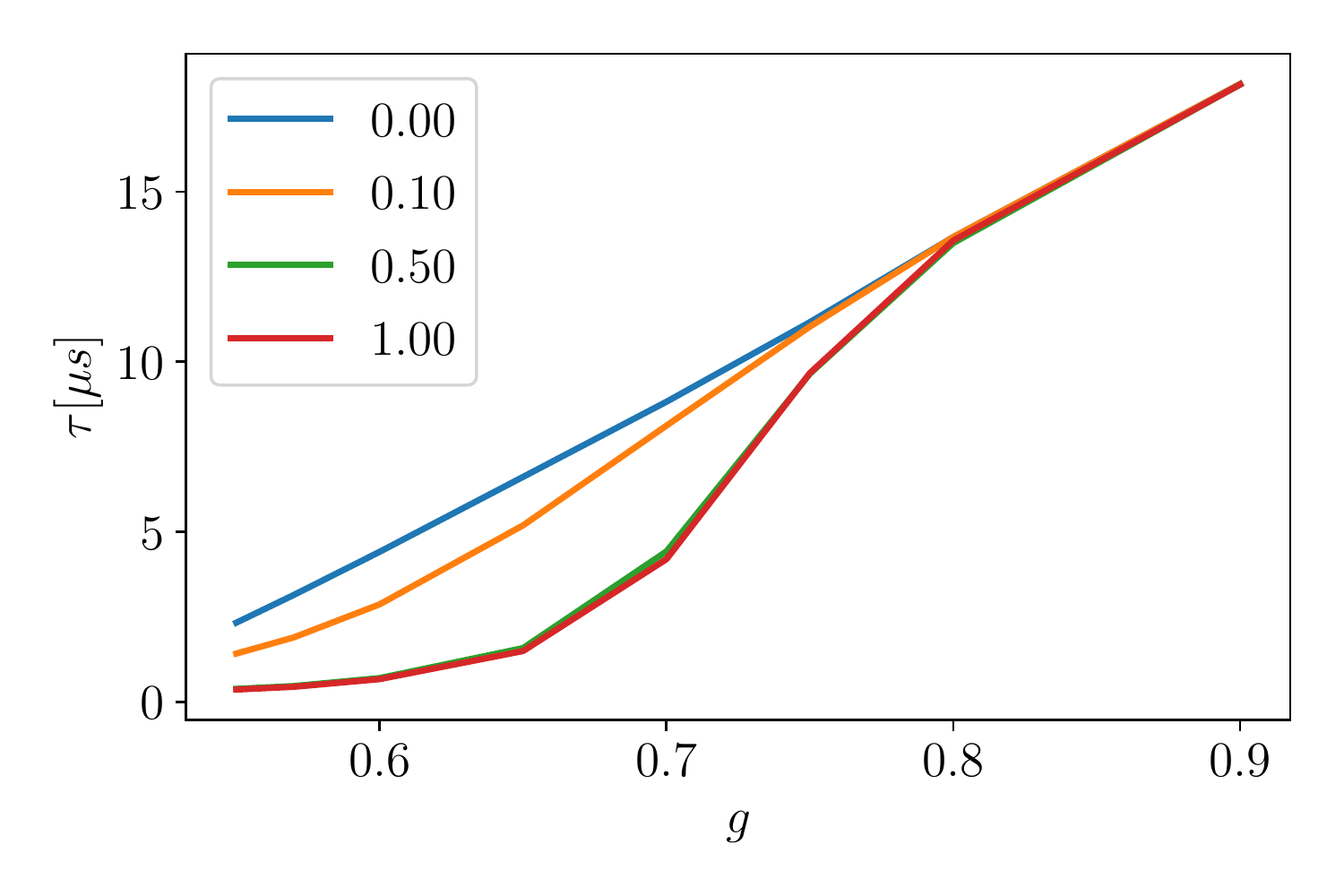} 
 \caption{%
Effect of integrability-breaking perturbation on the decay time of the Majorana operator, for $J=1/2$, as a function of $g$. As a perturbation, a kick with a uniform magnetic field $e^{-i  \sum_i h \hat Z_i}$ during each Floquet period was chosen. Inset illustrates the strength of the field $h$. We observe that the perturbation does not affect the decay time at $g\geq 0.8$, and reduce it at smaller values of $g$. Dephasing and decay parameters were chosen to be $\gamma_\phi=0.01, \gamma_d=0.0046$. 
 }
 \label{fig:heffect}
\end{figure}

\section{Simulations of dissipative many-body dynamics}\label{app:numerics}

To compute the temporal correlation function of local operators $\hat{O}$ we employ the influence matrix (IM) method \cite{LerosePRX,SonnerAoP}. We extended this method to treat open quantum systems. To that end, we start with the expression for the temporal correlation function in an open quantum system with discrete time
\begin{equation}
    \langle \hat{O}(t) \hat{O}(0) \rangle = \text{Tr}\left(\hat{O}\mathbf{C}^t[\hat{O}\hat{\rho}_0]\right)\label{eq:autocorrelator}
\end{equation}
where $\mathbf{C}[\cdot]$ is the channel corresponding to one step in time evolution. For our model, $\mathbf{C}[\cdot]$ consists of the unitary time evolution set by the Floquet operator $\hat{U}_{\rm F}$ and local dissipation (dephasing and decay with parameters $\gamma_d,\gamma_\phi$ respectively). Assuming that dissipation is weak, we approximate this channel by a subsequent application of $F$ and a purely dissipative quantum channel: 
\begin{align}
    \mathbf{C}[\rho]=\left(\bigotimes_i\mathbf{D}_i\right)\left[\hat{U}_{\rm F}\rho\hat{U}_{\rm F}^\dagger\right] && D_i\left[\left(\begin{array}{cc}\rho_{\uparrow\uparrow} & \rho_{\uparrow\downarrow}\\\rho_{\downarrow\uparrow}&\rho_{\downarrow\downarrow}\end{array}\right)\right] = \left(\begin{array}{cc}
        (1-\gamma_d)\rho_{\uparrow\uparrow} & (1-\gamma_\phi) \sqrt{1-\gamma_d} \rho_{\uparrow\downarrow} \\
         (1-\gamma_\phi)\sqrt{1-\gamma_d} \rho_{\downarrow\uparrow}& \rho_{\downarrow\downarrow} + \gamma_d \rho_{\uparrow\uparrow}
    \end{array}\right) 
\end{align}
Note that in the above equation we assumed the decay and dephasing parameters to be identical for all qubits, but the method can be straightforwardly adapted to account for non-uniform dissipation. 

The right-hand side of Eq. \eqref{eq:autocorrelator} can be understood as a tensor network (see Ref.~\cite{LerosePRX} for details). Calculating the temporal correlation functions reduces to contracting this tensor network. The strict light-cone in this tensor networks ensures that everything outside the light cone can be erased. We now contract this reduced tensor network from left to right until we reach the site where the operator $\hat{O}$ is localized. At each step we compress the tensor network to a matrix product state (MPS) with finite bond dimension $\chi=128,192$ by truncating the smallest singular values. This compression becomes more efficient as the temporal entanglement is lower. Interestingly, dissipation present in our model reduces the amount of temporal entanglement, making our approach more efficient. 

Once the influence matrix of the final time is obtained, calculating temporal correlations becomes straightforward: The individual tensors of the IM MPS can be interpreted as channels which act on the boundary spin (physical legs) as well as on a compressed representation of the quantum memory of the environment (virtual legs). Thus any correlation function can be computed by subsequently applying local time evolution and IM MPS tensors to an initial density matrix of a few spins.
To ensure that the calculations are not strongly affected by truncation to finite bond dimension we ran each simulation with maximal bond dimension $\chi=128$, as well as bond dimension $\chi=192$, and verified that they have converged.

The results of computations using IM method are illustrated in Figs.~\ref{fig:corr_decay},~\ref{fig:fit_rates},~\ref{fig:heffect}. In particular, in Fig.~\ref{fig:corr_decay} we illustrate temporal decay of multi-spin correlators that enter the Majorana operator expansion, and find that their decay rates are identical. Further, Fig.~\ref{fig:fit_rates} describes fits of the experimentally extracted decay times of long-lived correlators (see Fig.~\ref{fig:s8}B). Finally, Fig.~\ref{fig:heffect} illustrates the effect of integrability-breaking perturbation on the lifetime of the Majorana operators.  

\end{document}